\documentclass[prl,reprint, twocolumn,showpacs,preprintnumbers,amsmath,amssymb,nofootinbib,floatfix]{revtex4-1} 
\usepackage{graphicx}

\usepackage{mathrsfs}
\usepackage{hyperref}

\usepackage{slashed}

\usepackage{color}

\usepackage{gensymb}
\allowdisplaybreaks

\def \matrix #1 {\left(\begin{array}{cc} #1 \end{array}\right)}

\def\II{\hbox{{1}\kern-.25em\hbox{l}}}

\definecolor{mygray}{gray}{0.3}
\definecolor{HumboldtGreen}{rgb}{0,.4,0}

\begin{document}

\title{Shedding New Light on ${\cal R} (D_{(s)}^{(\ast)} )$ and $|V_{cb}|$ from
Semileptonic $\bar B_{(s)} \to D_{(s)}^{(\ast)} \ell \bar {\nu}_{\ell}$ Decays}

\author{Bo-Yan Cui}
\email{boyancui@nankai.edu.cn}

\author{Yong-Kang Huang}
\email{huangyongkang@mail.nankai.edu.cn}

\author{Yu-Ming Wang}
\email{correspondence author: wangyuming@nankai.edu.cn}

\author{Xue-Chen Zhao}
\email{correspondence author: zxc@mail.nankai.edu.cn}

\affiliation{\vspace{0.2 cm}
 School of Physics, Nankai University, Weijin Road 94, Tianjin 300071, P.R. China
\vspace{0.2 cm}}

\date{\today}

\begin{abstract}
\noindent

We compute for the first time the next-to-leading-order QCD corrections  to the
$\bar B_{(s)} \to D_{(s)}^{(\ast)}$  form factors at large hadronic recoil.
Both the charm-quark-mass and the strange-quark-mass  dependent pieces
can generate the leading-power contributions to these form factors.
Including  further various power-suppressed contributions,
we perform the combined fits of the considered form factors
to both our large-recoil theory predictions and the  lattice  QCD results,
thus improving upon the previous determinations of the lepton-flavour-universality ratios
${\cal R} (D^{(\ast)} )$ significantly.
\\[0.4em]

\end{abstract}


\maketitle

%
\section{Introduction}
%
%
The flagship semileptonic $\bar B_{(s)} \to D_{(s)}^{(\ast)} \ell \bar {\nu}_{\ell}$ decay processes
are of extraordinary phenomenological importance  for the precise determination of the
Cabibbo-Kobayashi-Maskawa (CKM) matrix element $|V_{cb}|$ 
and for probing the non-standard flavour-changing dynamics above the electroweak scale
\cite{Belle:2015pkj,Belle:2017rcc,Belle:2018ezy,LHCb:2020cyw,LHCb:2020hpv,Belle:2023bwv}.
The persisting discrepancy between the exclusive and inclusive extractions of  $|V_{cb}|$
at the level of  $2\, \sigma \sim 3 \, \sigma$ has triggered enormous efforts
on uncovering the ultimate mystery of this intriguing puzzle in the Standard Model (SM) and beyond
(see for instance \cite{Belle-II:2018jsg,Gambino:2020jvv,Boyle:2022uba,USQCD:2022mmc,DiCanto:2022icc}).
Moreover, the state-of-the-art theory predictions for the two lepton-flavour-universality (LFU) ratios
${\cal R}({D^{(\ast)}})$ \cite{Bernlochner:2021vlv}
appear to be in tension with the HFLAV-averaged experimental measurements  \cite{HFLAV:2022pwe}
(at the  $3 \, \sigma$ level when combined together) as well,
thus stimulating intensive investigations on a wide range of
other interesting  LFU observables \cite{Bernlochner:2016bci,Bernlochner:2018kxh,Cohen:2018vhw,Das:2021lws,Gubernari:2022hrq,
Bernlochner:2022hyz,Patnaik:2022moy,Patnaik:2023ins,Penalva:2023snz}.
Apparently,  disentangling the potential new physics (NP) signals
from the unaccounted theoretical uncertainties in the SM computations
in a robust manner  will be indispensable for the unambiguous interpretation of
these emerged flavour anomalies.
It remains important to remark that the updated LHCb measurements for the electron-muon universality rations ${\cal R} (K^{(\ast)})$
in the flavour-changing neutral current loop processes $B \to K^{(\ast)} e^{+} e^{-}$ and $B \to K^{(\ast)} \mu^{+} \mu^{-}$
\cite{LHCb:2022qnv,LHCb:2022zom} do not necessarily lead to conclude the tauon-muon universality in the
neutral current $b \to s \ell \bar \ell$ transitions \cite{SinghChundawat:2022ldm,Alok:2023yzg},
let alone in the flavour-changing charged current tree  decays
$\bar B_{(s)} \to D_{(s)}^{(\ast)} \ell \bar {\nu}_{\ell}$ \cite{Alguero:2023jeh}.
Actually, introducing new CP-violating couplings in the weak  effective Hamiltonian for $b \to s \ell \bar \ell$
can bring about the significant space to violate  the electron-muon universality,
while accommodating the new LHCb result for the ${\cal R} (K)$ ratio simultaneously \cite{Fleischer:2023zeo}.

The model-independent descriptions of the heavy-to-heavy bottom-meson decay form factors
in the low recoil regime can be naturally formulated  by adopting the heavy quark effective theory (HQET)
based upon an expansion in powers of $\Lambda_{\rm QCD}/m_{b, \, c}$,
yielding a tower of the non-perturbative Isgur-Wise (IW) functions.
In order to better constrain the desirable shape of these form factors,
an attractive prescription to derive the  unitarity constraints
for the form-factor expansion coefficients has been developed from
the fundamental field-theoretical principles  
\cite{Boyd:1995cf,Boyd:1995sq,Boyd:1997kz,Caprini:1996vw,Caprini:1997mu,Bourrely:2008za}.
Additionally, we have  witnessed the substantial progress
on the encouraging lattice QCD calculations of the $\bar B \to D^{(\ast)} \ell \bar \nu_{\bar \ell}$ form factors
\cite{MILC:2015uhg,Na:2015kha,FermilabLattice:2021cdg}
and the $\bar B_s \to D_s^{(\ast)} \ell \bar \nu_{\bar \ell}$ form factors
\cite{Bailey:2012rr,McLean:2019qcx,Harrison:2021tol} at non-zero recoil.
By contrast, the continuum QCD determinations of such fundamental  heavy-to-heavy form factors at large recoil
are mainly achieved by evaluating only  the leading-order QCD contributions  at the twist-four accuracy
\cite{Faller:2008tr,Gubernari:2018wyi},  
apart from the currently available higher-order QCD computations
of the  $\bar B \to D$ form factors \cite{Wang:2017jow,Gao:2021sav}.

In view of the noticeable significance of implementing the large-recoil theory predictions
in the numerical fit of the exclusive bottom-meson decay form factors
\cite{Arnesen:2005ez,Cui:2022zwm,Bigi:2017jbd,Gambino:2019sif,Bordone:2019vic,
Bordone:2019guc,Jaiswal:2017rve,Jaiswal:2020wer,Biswas:2022yvh,Cheung:2020sbq},
accomplishing   the next-to-leading-order (NLO) computations to 
the semileptonic $\bar B_{(s)} \to D_{(s)}^{\ast} \ell \bar \nu_{\bar \ell}$ form factors
will therefore be in  high demand for pinning down the obtained uncertainties
of their shape parameters from the combined $z$-series fitting procedure.
To achieve this goal, we will first establish the NLO factorization formulae
for the appropriate bottom-meson-to-vacuum correlation functions at leading power (LP)
in the soft-collinear effective theory (SCET) framework
and then construct the large logarithmic resummation improved light-cone sum rules (LCSR)
for the considered form factors. 
In particular, we will report on a novel observation of the LP ${\cal O}(\alpha_s)$ contribution
to the longitudinal form factors due to the unsuppressed charm-quark mass dependent pieces in the SCET Lagrangian,
when applying the  preferable  power-counting scheme  $m_c \sim {\cal O} \left (\sqrt{\Lambda_{\rm QCD} \, m_b }\right)$
\cite{Boos:2005by,Boos:2005qx}.
Phenomenological implications of the simultaneous Boyd-Grinstein-Lebed (BGL) expansion fitting
\cite{Boyd:1995cf,Boyd:1995sq,Boyd:1997kz} to both the SCET sum rules predictions and the lattice simulation data points
will be further explored with the focus on the updated extractions of the LFU quantities
${\cal R} (D_{(s)}^{(\ast)} )$ and the CKM matrix element $|V_{cb}|$.

%
\section{General analysis}
%

\begin{widetext}
We adopt the customary definitions of the  bottom-meson decay form factors
$\{V, \, A_0, \, A_1, \, A_{2} \}$ as displayed in \cite{Gao:2019lta}.
Implementing the  matching program  ${\rm QCD} \to {\rm SCET_{I}}$
for these form factors enables us to derive the  factorization formulae
in terms of the  SCET form factors
\begin{eqnarray}
{\cal V}(n \cdot p)
&=&  C_V^{(\rm A0)} \left (n \cdot p \right )  \xi_{\perp}(n \cdot p)
+ C_V^{({\rm A1}, \, m_c)} \, \left (n \cdot p \right ) \, \xi_{\perp, \,  m_c}(n \cdot p)
+ \int_0^1  d \tau \, C_V^{(\rm B1)}
\left (\tau, n \cdot p  \right )  \Xi_{\perp}(\tau, n \cdot p)
+ {\cal V}^{\rm NLP}(n \cdot p) \,,
\hspace{0.5 cm}
\nonumber \\
{\cal A}_0(n \cdot p)
&=&  C_{f_0}^{(\rm A0)} \left (n \cdot p \right )  \xi_{\|}(n \cdot p)
+ C_{f_0}^{({\rm A1}, \, m_c)} \, \left (n \cdot p \right ) \, \xi_{\|, \,  m_c}(n \cdot p)
+ \int_0^1  d \tau \, C_{f_0}^{(\rm B1)}
\left (\tau, n \cdot p  \right )  \Xi_{\|}(\tau, n \cdot p)
+ {\cal A}_{0}^{\rm NLP}(n \cdot p),
\hspace{0.5 cm}
\nonumber \\
{\cal A}_1(n \cdot p)
&=&  C_V^{(\rm A0)} \left (n \cdot p \right )  \xi_{\perp}(n \cdot p)
+ C_{A_1}^{({\rm A1}, \, m_c)} \, \left (n \cdot p \right ) \, \xi_{\perp, \,  m_c}(n \cdot p)
+ \int_0^1  d \tau \, C_V^{(\rm B1)}
\left (\tau, n \cdot p  \right )  \Xi_{\perp}(\tau, n \cdot p)
+ {\cal A}_{1}^{\rm NLP}(n \cdot p) \,,
\hspace{0.5 cm}
\nonumber \\
{\cal A}_{12}(n \cdot p)
&=&  C_{f_+}^{(\rm A0)} \left (n \cdot p \right )  \xi_{\|}(n \cdot p)
+ C_{f_+}^{({\rm A1}, \, m_c)} \, \left (n \cdot p \right ) \, \xi_{\|, \,  m_c}(n \cdot p)
+ \int_0^1  d \tau \, C_{f_+}^{(\rm B1)}
\left (\tau, n \cdot p  \right )  \Xi_{\|}(\tau, n \cdot p)
+ {\cal A}_{12}^{\rm NLP}(n \cdot p).
\hspace{0.5 cm}
\label{factorization of FFs: mater formulae}
\end{eqnarray}
\end{widetext}
The newly introduced form factors ${\cal V}$, ${\cal A}_{0}$, ${\cal A}_{1}$ and ${\cal A}_{12}$ can be expressed
in terms of the linear combinations of the conventional  form factors as displayed in (1) of
the Supplemental Material.
The explicit definitions of $\xi_{\|, \, \perp}$ and $\Xi_{\|, \, \perp}$
take the same form as the ones for the exclusive heavy-to-light transitions \cite{Beneke:2005gs},
while the remaining two effective form factors  $\xi_{\|, \,  m_c}$ and $\xi_{\perp, \,  m_c}$
can be defined by
\begin{eqnarray}
&& \langle  D_q^{\ast} (p, \epsilon^{\ast}) | (\bar \xi W_c) \,  {\slashed{n} \over 2}
{m_c  \over - i n \cdot \overleftarrow{D}_c} \gamma_5 \,   h_v | \bar B_v  \rangle
\nonumber \\
&& = - n \cdot p  \, \left ( \epsilon^{\ast} \cdot v \right ) \,  \xi_{\|, \,  m_c}(n \cdot p),
\label{definition: NLP SCET longitudinal FF}
\\
&& \langle  D_q^{\ast} (p, \epsilon^{\ast}) |  (\bar \xi W_c) \,  {\slashed{n} \over 2}
{m_c  \over - i n \cdot \overleftarrow{D}_c} \gamma_5 \gamma_{\mu \perp} \,  h_v | \bar B_v  \rangle
\nonumber \\
&& = - n \cdot p \,  \left (\epsilon^{\ast}_{\mu} - \epsilon^{\ast} \cdot v \, \bar n_{\mu}\right )  \,
\xi_{\perp, \,  m_c}(n \cdot p) \,.
\label{definition: NLP SCET transverse FF}
\end{eqnarray}
The analytic expressions for the short-distance coefficients $C_{i}^{(\rm A0)}$ and $C_{i}^{(\rm B1)}$
(determined from \cite{Beneke:2004rc,Hill:2004if}) as well as   $C_{i}^{({\rm A1}, \, m_c)}$
are collected in (2) of the Supplemental Material.
We will therefore dedicate the next section to the transparent computations of
the  effective form factors at ${\cal O}(\alpha_s)$,
including further the tree-level determinations of the next-to-leading power (NLP) corrections.

%
\section{NLO corrections to the form factors}
\label{section: NLO sum rules}
%

In analogy to the  strategy for computing  the heavy-to-light bottom-meson decay matrix elements
\cite{DeFazio:2005dx,DeFazio:2007hw} (see also \cite{Khodjamirian:2005ea,Khodjamirian:2006st}),
we can derive the  LCSR for  $\xi_{\|}(n \cdot p)$
by exploring  the particular ${\rm SCET_I}$ correlation function
\begin{eqnarray}
&& \Pi_{\nu, \, \|}^{\rm (A0)}
= \int d^4 x \, e^{i p \cdot x} \,
\langle  0 | {\rm T} \{j_{\xi q, \| \nu}^{(2)}(x), \,\,  O_{\|}^{\rm (A0)}(0)  \}  | \bar B_v \rangle
\nonumber  \\
&&  +  \int d^4 x \, e^{i p \cdot x} \, \int d^4 y  \,
\nonumber \\
&& \hspace{0.3 cm} \langle  0 | {\rm T} \{j_{\xi \xi, \| \nu}^{(0)}(x), \,\, i {\cal L}_{\xi q}^{(2)}(y),
\,\,  O_{\|}^{\rm (A0)}(0)  \}  | \bar B_v \rangle
\nonumber \\
&& +  \int d^4 x \, e^{i p \cdot x} \, \int d^4 y  \, \int d^4 z  \,
\nonumber \\
&& \hspace{0.3 cm} \langle  0 | {\rm T} \{j_{\xi \xi, \| \nu}^{(0)}(x), \,\, i {\cal L}_{\xi q}^{(1)}(y),
\,\, i {\cal L}_{\xi m_c}^{(0)}(z), \,\,  O_{\|}^{\rm (A0)}(0)  \}  | \bar B_v \rangle
\nonumber \\
&& +  \int d^4 x \, e^{i p \cdot x} \, \int d^4 y  \,
\nonumber \\
&& \hspace{0.3 cm} \langle  0 | {\rm T} \{j_{\xi \xi, \| \nu}^{(0)}(x), \,\, i {\cal L}_{\xi q, m_q}^{(2)}(y),
\,\,  O_{\|}^{\rm (A0)}(0)  \}  | \bar B_v \rangle     \,,
\label{the longitudinal correlation function: A0}
\end{eqnarray}
where the manifest representations of
$j_{\xi \xi, \| \nu}^{(0)}$, $j_{\xi q, \| \nu}^{(2)}$, ${\cal L}_{\xi q}^{(1)}$
and ${\cal L}_{\xi q}^{(2)}$ have been presented in \cite{Beneke:2002ni,Gao:2019lta}.
The remaining effective Lagrangian densities and the ${\rm SCET_I}$ weak current
are presented in (3) of the Supplemental Material.
Since the charm-quark mass dependent term ${\cal L}_{\xi m_c}^{(0)}$
describes the unsuppressed interaction between the collinear fields \cite{Leibovich:2003jd},
the third term in the correlation function (\ref{the longitudinal correlation function: A0})
will result in  the power-enhanced contribution to
{\it the correlation function} (\ref{the longitudinal correlation function: A0}).
Moreover, the spectator-quark mass contributions from  the second  and the fourth terms
on the right-hand side of (\ref{the longitudinal correlation function: A0})
can further give rise to the LP effects
(see also \cite{Leibovich:2003jd,Boer:2018mgl,Cui:2022zwm}).

Matching the determined spectral representation of $\Pi_{\nu, \, \|}^{\rm (A0)}$
with the corresponding hadronic dispersion relation leads to the desired NLO sum rules
\begin{eqnarray}
\xi_{\|} &=&  2 \, \frac{{\cal F}_{B_q}(\mu)}{f_{D_q^{\ast}, \|}} \,
{m_{B_q} m_{D_q^{\ast}} \over (n \cdot p)^2} \, \int_0^{\omega_s} d \omega^{\prime} \,
{\rm exp} \left [  {m_{D_q^{\ast}} ^2 - n \cdot p \, \omega^{\prime}    \over n \cdot p \, \omega_M}  \right ]
\nonumber \\
&&  \left [ \phi_{B, \, {\rm eff}}^{-}(\omega^{\prime})
- { m_c \over \omega^{\prime}} \, \phi_{B, \,  {\rm eff}}^{+, \, m_c}(\omega^{\prime})
+  { m_q \over \omega^{\prime}} \, \phi_{B, \,  {\rm eff}}^{+, \, m_q}(\omega^{\prime})  \right ]
\nonumber \\
&\equiv&  \hat{\xi}_{\|} + \hat{\xi}_{\|}^{\, m_c} + \hat{\xi}_{\|}^{\, m_q}   \,,
\label{effective form factor for xi_L}
\end{eqnarray}
where the two decay constants ${\cal F}_{B_q}$  and  $f_{D_q^{\ast}, \|}$
are defined with the  conventions of \cite{Gao:2019lta}.
The lengthy expressions for $\phi_{B, \, {\rm eff}}^{-}$,
$\phi_{B, \,  {\rm eff}}^{+, \, m_c}$ and $\phi_{B, \,  {\rm eff}}^{+, \, m_q}$
are summarized in  the Supplemental Material.
Remarkably, the ${\rm SCET_I}$ diagram (b) in Figure 1 of the Supplemental Material
does not lead to the power-enhanced contribution (but does generate the LP contribution)
to the  sum rules of $\xi_{\|}$ after implementing {\it  the continuum subtraction}.
Applying the established computational strategy further allows for the construction of  the SCET sum rules
for the effective  form factors $\xi_{\|, \,  m_c}$ and $\Xi_{\|}$
(shown in (8) and (12) of the Supplemental Material)
by investigating the appropriate  correlation functions.
Along the same vein, we can readily derive the NLO sum rules for the transverse form factors
$\xi_{\perp}$, $\xi_{\perp, \,  m_c}$ and $\Xi_{\perp}$ as collected in the Supplemental Material.

We then proceed to  construct the subleading-power sum rules for the
$\bar B_q \to D_q^{\ast}$ form factors from four distinct sources:
I) the off-light-cone corrections from the two-body nonlocal HQET matrix elements,
II) the yet higher-twist corrections from the
three-particle $B_q$-meson distribution amplitudes
\cite{Geyer:2005fb,Braun:2017liq},
III) the subleading-power corrections from the effective matrix element
of  $(\bar \xi W_c) \, \Gamma \, [i \, \slashed{D}_{\rm \top} / (2 m_b)] \, h_v$ \cite{Beneke:2002ni},
IV) the higher-order terms from expanding the hard-collinear charm-quark propagator.
The tree-level LCSR for the power-suppressed terms ${\cal V}^{\rm NLP}$, ${\cal A}_0^{\rm NLP}$,
${\cal A}_1^{\rm NLP}$ and ${\cal A}_{12}^{\rm NLP}$ will be presented explicitly in a forthcoming longer write-up.

%
\section{Numerical analysis}
%

We are now in a position to explore phenomenological implications of
the newly determined SCET sum rules
with the three-parameter ans\"{a}tz for the  bottom-meson distribution amplitudes
\cite{Beneke:2018wjp}
(see also \cite{Bell:2013tfa,Wang:2015vgv,Wang:2016qii,Wang:2018wfj,Shen:2020hfq,Wang:2021yrr,Feldmann:2022uok}),
which fulfills  simultaneously the non-trivial equations-of-motion constraints \cite{Beneke:2018wjp},
the sum rule determinations of the two inverse moments $\lambda_{B_{d, s}}$ \cite{Braun:2003wx,Khodjamirian:2020hob}
and  the HQET parameters $\lambda_{E, \, H}^2$ \cite{Grozin:1996pq,Nishikawa:2011qk,Rahimi:2020zzo},
and the asymptotic behaviours at small quark and gluon momenta \cite{Braun:2017liq}.
It is perhaps worth mentioning an attractive method for the first-principles determination of
the $B$-meson light-cone distribution amplitude based upon the large momentum effective
theory and the lattice simulation technique \cite{Wang:2019msf}.
The leptonic decay constants of the pseudoscalar bottom mesons have been  extracted
from  the lattice calculations precisely \cite{FlavourLatticeAveragingGroupFLAG:2021npn}.
We  further 
employ the  QCD sum rule computations \cite{Pullin:2021ebn}
for both  the longitudinal and transverse $D_{q}^{\ast}$ decay constants.
The allowed intervals of the  intrinsic LCSR parameters
$M^2 = n \cdot p \, \omega_M$ and $s_0=n \cdot p \, \omega_s$
are consistent with the previous  determinations
\cite{Khodjamirian:2009ys,Duplancic:2015zna,Li:2020rcg,Khodjamirian:2020mlb}.
The choices for the additional  parameters in our numerical studies
are identical to the ones summarized in \cite{Cui:2022zwm}.

\begin{figure}[h]
\begin{center}
\includegraphics[width=0.90 \columnwidth]{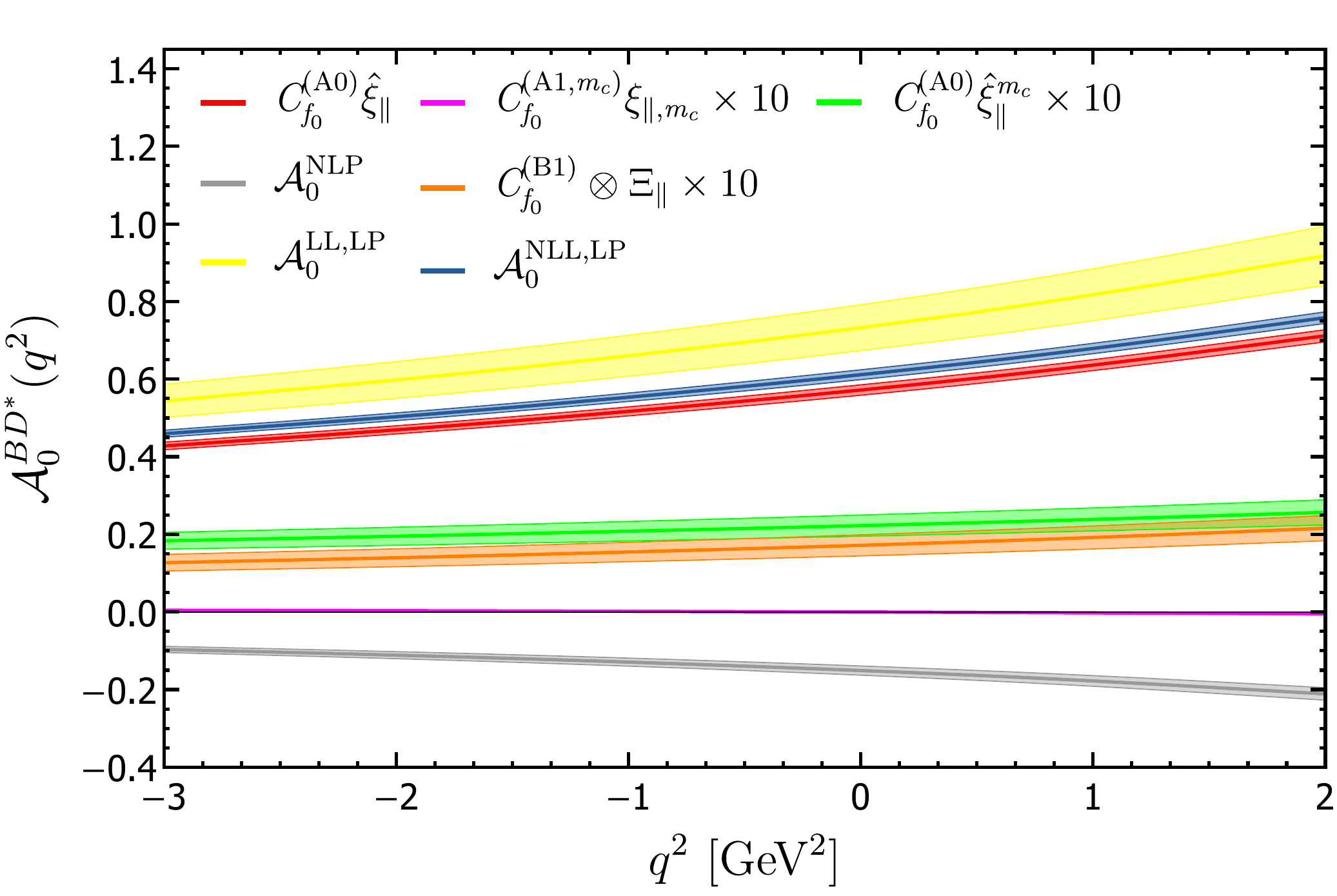}
\hspace{1.0 cm}
\caption{Breakdown of the distinct dynamical mechanisms contributing to the  form factor ${\cal A}_0$
for  $\bar B \to D^{\ast} \, \ell \, \bar {\nu}_{\ell}$  in the kinematic  range
$q^2 \in [-3.0, \, 2.0] \, {\rm GeV^2}$ with the  uncertainties from varying the hard and hard-collinear
matching scales.  }
\label{fig: Breakdown of the B to Dstar form factor A0}
\end{center}
\end{figure}

\begin{table*}[tp]
\centering
\renewcommand{\arraystretch}{1.0}
\resizebox{2.0 \columnwidth}{!}{
\begin{tabular}{|c||ccc||ccc|}
\hline
\hline
  & \multicolumn{3}{|c||}{$\bar B \to D^{(\ast)}$ \, Form Factors}  & \multicolumn{3}{|c|}{$\bar B_s \to D_s^{(\ast)}$ \, Form Factors}
\\
\hline
Parameters
& Lattice
& Lattice $\oplus$ LCSR
& Lattice $\oplus$ LCSR $\oplus$ Exp.
& Lattice
& Lattice $\oplus$ LCSR
& Lattice $\oplus$ LCSR $\oplus$ Exp.
\\
\hline
\hline
  $b_0^{f_{+}}$             &   $\ \ \,0.0137 \pm 0.0001$      & $\ \ \,0.0137 \pm 0.0001$
  &   $\ \ \,0.0138 \pm 0.0001 $  &   $\ \ \,0.0041 \pm 0.0001$      & $\ \ \,0.0041 \pm 0.0001$              &  $\ \ \,0.0042 \pm 0.0001$
\\
  $b_1^{f_{+}}$             &   $-0.0414 \pm 0.0034$     &  $\,\, -0.0417 \pm 0.0033\,\,$
  &  $-0.0398 \pm 0.0032$   &   $-0.0029 \pm 0.0019$    &  $\,\,-0.0030 \pm 0.0019\,\,$     &  $-0.0034 \pm  0.0015$
\\
  $b_2^{f_{+}}$             &   $\ \ \,0.1178 \pm 0.2007$      & $\ \ \,0.0415 \pm 0.1124 $
  &  $\ \ \,0.1123 \pm 0.0713 $   &   $-0.0584 \pm 0.0096$     & $ -0.0588 \pm 0.0093$            &  $-0.0608 \pm 0.0078$
\\
  $b_1^{f_{0}}$             &   $-0.2064 \pm 0.0155 $     & $-0.2072 \pm 0.0147$
  &  $-0.2004 \pm 0.0144$   &   $-0.0610 \pm 0.0112$        &  $-0.0617 \pm 0.0111$             &  $-0.0571 \pm 0.0108$
\\
  $b_2^{f_{0}}$             &   $\ \ \,0.5572 \pm 0.9626$      & $\ \ \,0.1880 \pm 0.5330$
  &  $\ \ \,0.5581 \pm 0.3297$      &   $-0.0264 \pm 0.0768 $     &  $-0.0233 \pm 0.0767 $             &  $-0.0359 \pm 0.0756$
\\
\hline
\hline
  $b_0^{g}$                 &    $\ \ \,0.0259 \pm 0.0009 $    & $\ \ \,0.0256 \pm 0.0009$
&  $\ \ \,0.0251 \pm  0.0008 $    &    $\ \ \,0.0080 \pm 0.0009 $    &  $\ \ \,0.0072 \pm 0.0007$             &  $\ \ \,0.0071 \pm 0.0007$
\\
  $b_1^{g}$                 &   $-0.1093 \pm 0.0786$     & $-0.1005 \pm 0.0456$
&  $-0.1104 \pm 0.0396$       &   $\ \ \, 0.0212 \pm 0.0218$     & $ -0.0017 \pm 0.0106$            &  $-0.0005 \pm 0.0105$
\\
  $b_2^{g}$                 &   $-0.4505 \pm 4.3691$     & $\ \ \,0.2587 \pm 0.6564$
&  $\ \ \,0.2050 \pm 0.6107$        &   $-0.0089 \pm 0.1334$     & $-0.1067 \pm 0.0848$             &  $-0.0964 \pm 0.0841$
\\
  $b_0^{f}$                 &   $\ \ \,0.0108 \pm 0.0002$      & $\ \ \,0.0109 \pm 0.0002 $
&  $\ \ \,0.0106 \pm 0.0002 $     &   $\ \ \,0.0035 \pm 0.0001$      & $\ \ \, 0.0036 \pm 0.0001$             &  $\ \ \,0.0036 \pm  0.0001$
\\
  $b_1^{f}$                 &   $-0.0012 \pm 0.0168$     & $\ \ \,0.0081 \pm 0.0101$
&  $\ \ \,0.0112 \pm 0.0082$      &   $\ \ \,0.0041 \pm 0.0036$      &  $\ \ \,0.0060 \pm 0.0034$             &   $\ \ \,0.0059 \pm 0.0032$
\\
  $b_2^{f}$                 &    $-0.0379 \pm 1.1507$     & $\ \ \,0.0693 \pm 0.2140$
& $-0.0713 \pm 0.1583$        &    $-0.0055 \pm 0.0531$     &  $\ \ \,0.0287 \pm 0.0447$            &  $\ \ \,0.0419 \pm 0.0437$
\\
  $b_1^{F_1}$               &    $-0.0046 \pm 0.0031$     &  $-0.0024 \pm 0.0020$
&  $\ \ \,0.0015 \pm 0.0014 $     &    $\ \ \,0.0013 \pm 0.0009$     &   $\ \ \,0.0013 \pm 0.0008 $          &   $\ \ \,0.0019 \pm 0.0007$
\\
  $b_2^{F_1}$               &     $-0.0460 \pm 0.1944 $    &  $-0.0155 \pm 0.0388$
&  $-0.0089 \pm 0.0251$       &     $-0.0188 \pm 0.0156 $    &  $ -0.0221 \pm 0.0126 $          &   $-0.0177 \pm 0.0122$
\\
  $b_1^{F_2}$               &      $-0.2949 \pm 0.0742$    &  $-0.2097 \pm 0.0581$
& $-0.1364 \pm 0.0490$        &      $-0.0078 \pm 0.0190$    &  $-0.0093 \pm 0.0145$         &   $\ \ \,0.0001 \pm 0.0136$
\\
  $b_2^{F_2}$               &      $\ \ \,0.5476 \pm 4.0297 $    &  $\ \ \,0.5667 \pm 0.8789$
&  $\ \ \,0.5660 \pm 0.7406$        &      $-0.2242 \pm 0.1756$    &  $ -0.1906 \pm 0.1521$           &   $-0.1480 \pm 0.1495$
\\
\hline
\hline
  $\chi^2/{\rm dof}$        &    $7.71/8$                  &  $30.76/76$
&  $118.31/140$             &    $7.71/8$                  &   $30.76/76$                    &  $118.31/140$
\\
  $R(D_q)$                  &    $\ \ \,0.3004 \pm 0.0143$       &   $\ \ \,0.3066 \pm 0.0081$
&  $\ \ \,0.2986 \pm 0.0042$     &    $\ \ \,0.2993 \pm  0.0046$      &   $\ \ \,0.2996 \pm 0.0045$           &  $\ \ \,0.2971 \pm 0.0042$
\\
  $R(D_q^{\ast})$           &    $\ \ \,0.2718 \pm 0.0300$       &   $\ \ \,0.2585 \pm 0.0054$
&  $\ \ \,0.2500 \pm 0.0016$      &    $\ \ \,0.2488 \pm 0.0058$       &   $\ \ \,0.2506 \pm 0.0040$           &  $\ \ \,0.2461 \pm 0.0024$
\\
\hline
\hline
\end{tabular}
}
\renewcommand{\arraystretch}{1.0}
\caption{Theory predictions for the $z$-series expansion coefficients in the semileptonic
$\bar B_{(s)} \to D_{(s)}^{(\ast)} \, \ell \, \bar {\nu}_{\ell}$  form   factors
determined by carrying out the BGL fitting against the ``only lattice QCD" data points
\cite{MILC:2015uhg,Na:2015kha,FermilabLattice:2021cdg,McLean:2019qcx,Harrison:2021tol}
(shown in the second and the fifth columns), by performing the simultaneous fitting to both the lattice QCD results
\cite{MILC:2015uhg,Na:2015kha,FermilabLattice:2021cdg,McLean:2019qcx,Harrison:2021tol}
and the updated LCSR computations (shown in the third and the sixth columns),
and by further implementing  the available experimental  data points
\cite{Belle:2015pkj,Belle:2018ezy,LHCb:2020cyw,LHCb:2020hpv} in the combined BGL fit procedure
(shown in the fourth  and the last columns).}
\label{table: BGL fit results of the expansion coefficients}
\end{table*}

\begin{figure*}[tp]
\begin{center}
\includegraphics[width=0.65 \columnwidth]{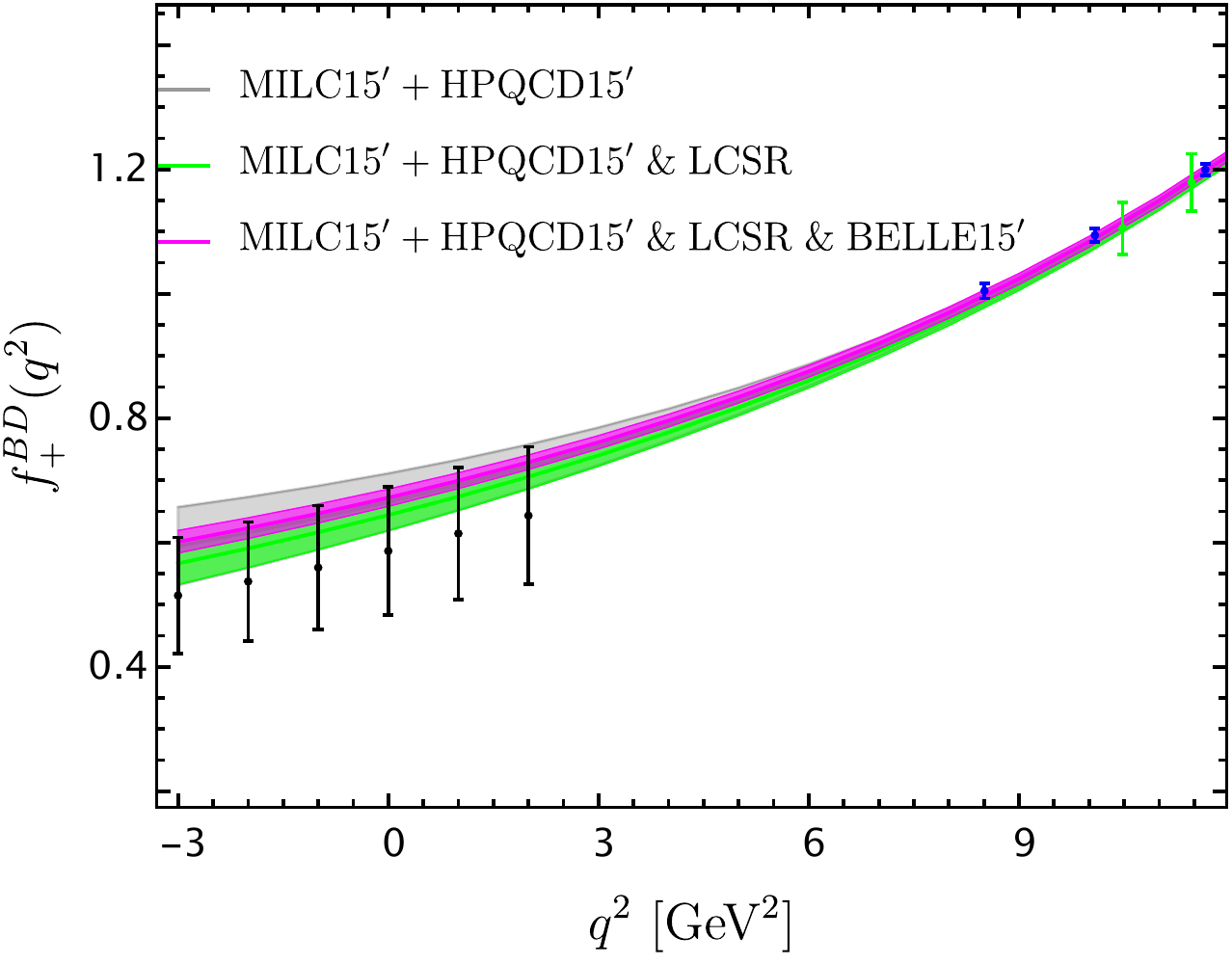}
\includegraphics[width=0.65 \columnwidth]{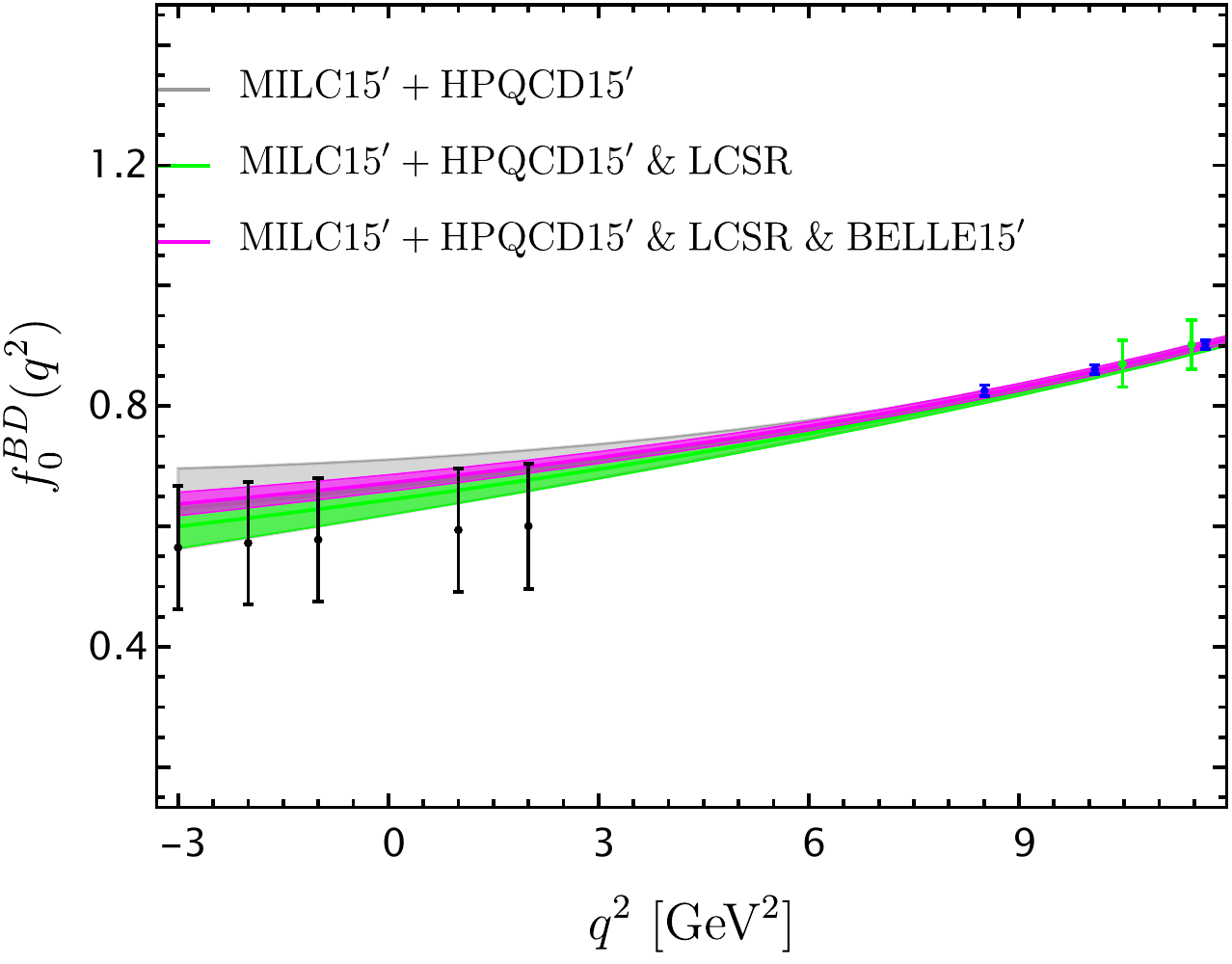}
\includegraphics[width=0.65 \columnwidth]{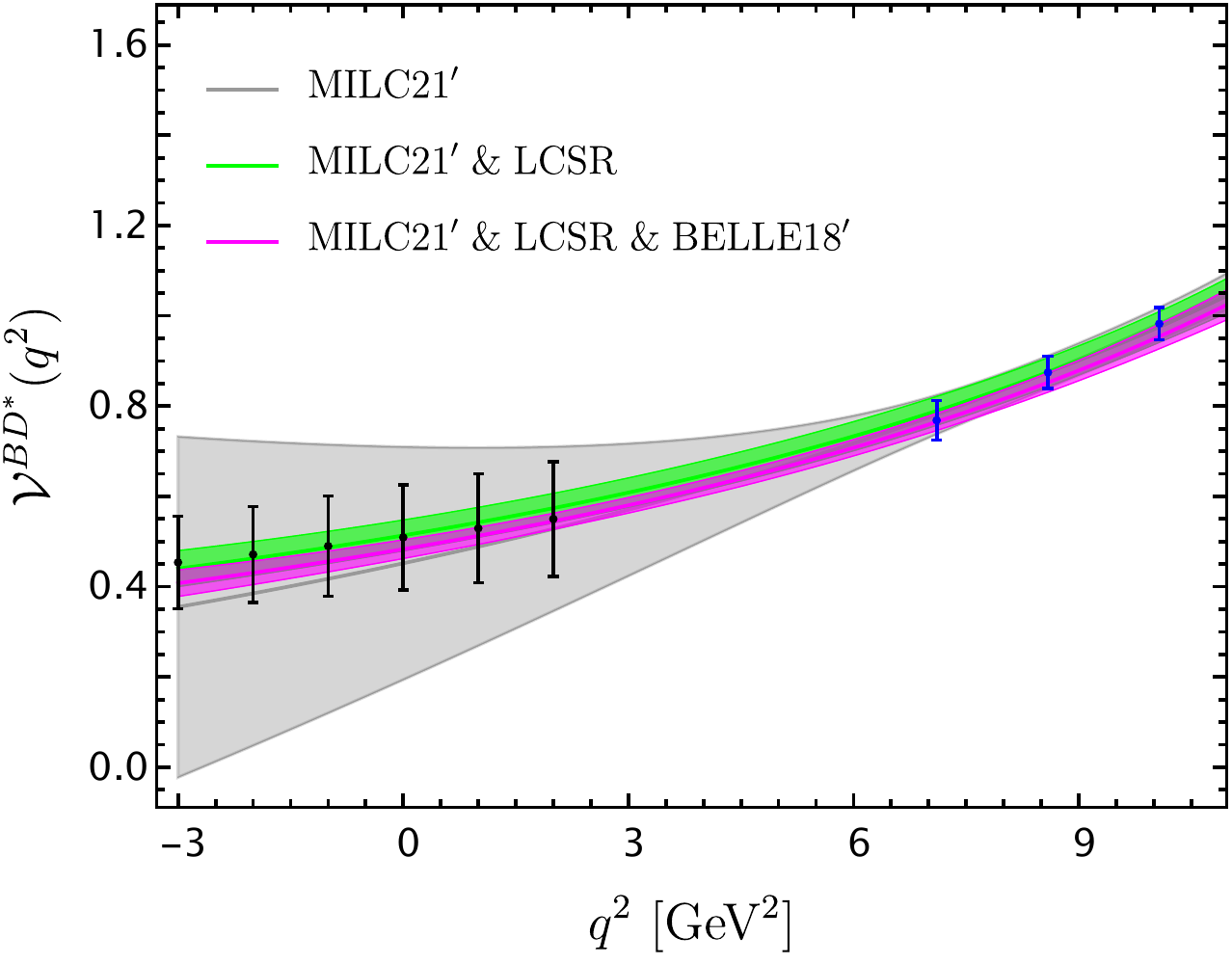}
\\
\includegraphics[width=0.65 \columnwidth]{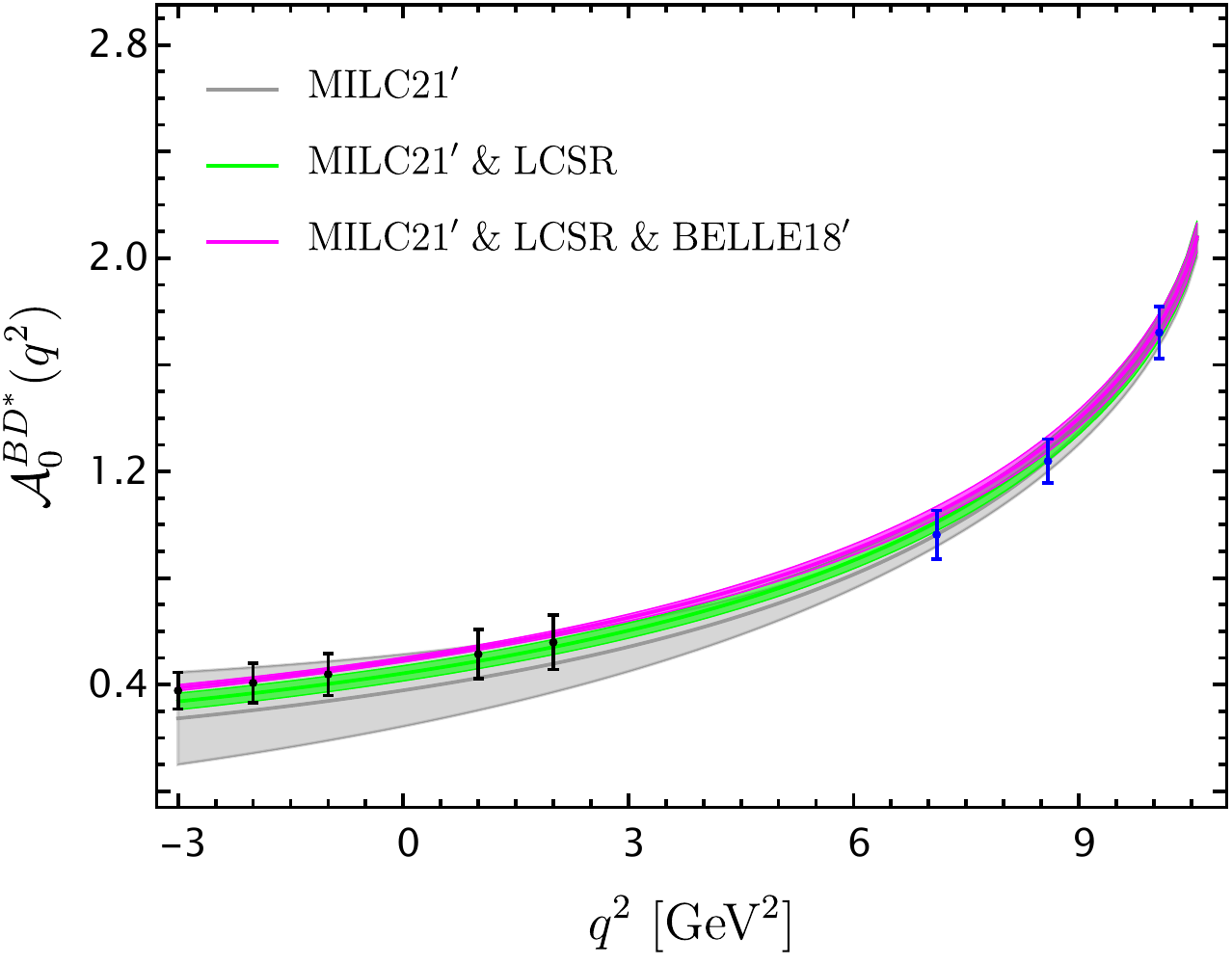}
\includegraphics[width=0.65 \columnwidth]{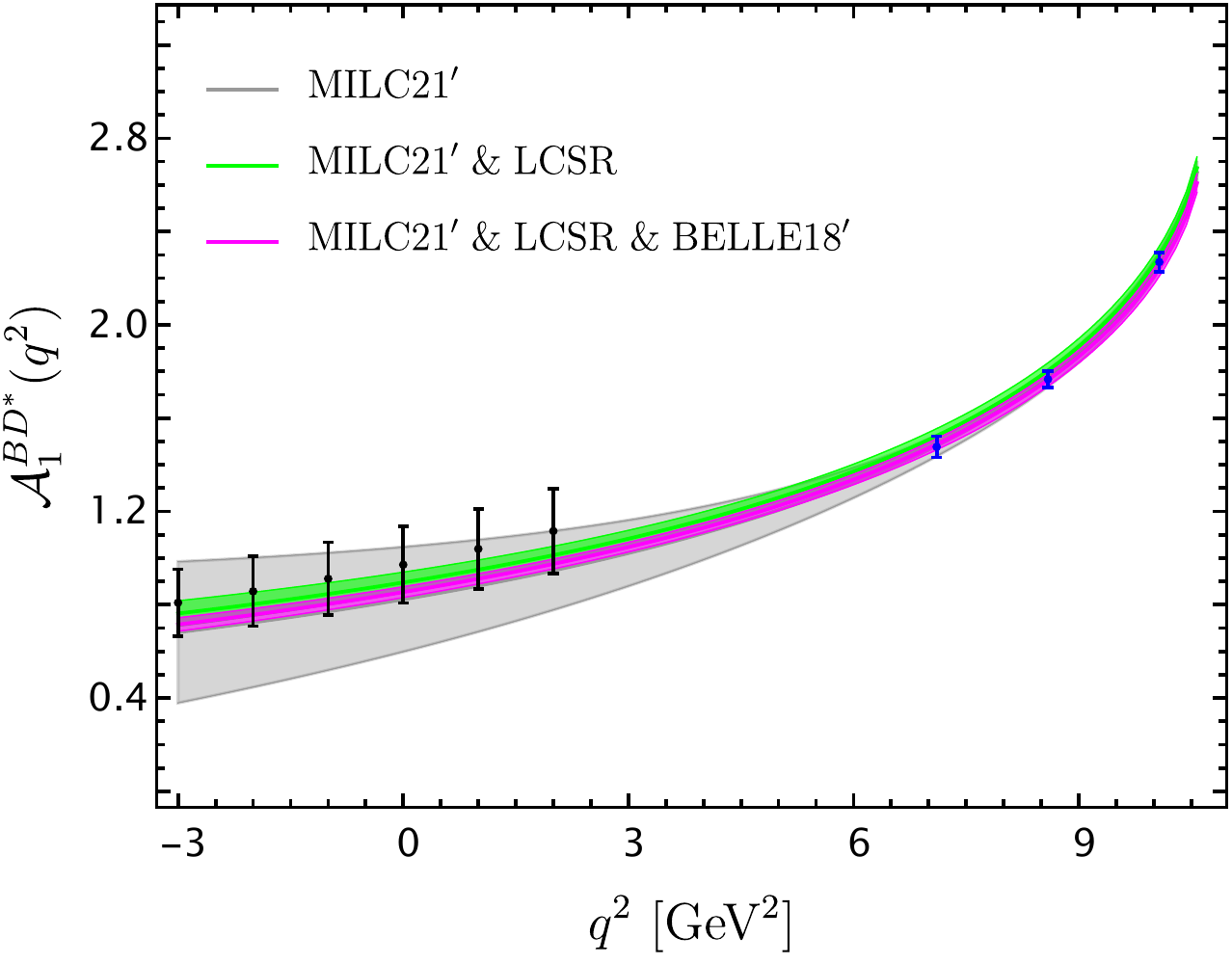}
\includegraphics[width=0.65 \columnwidth]{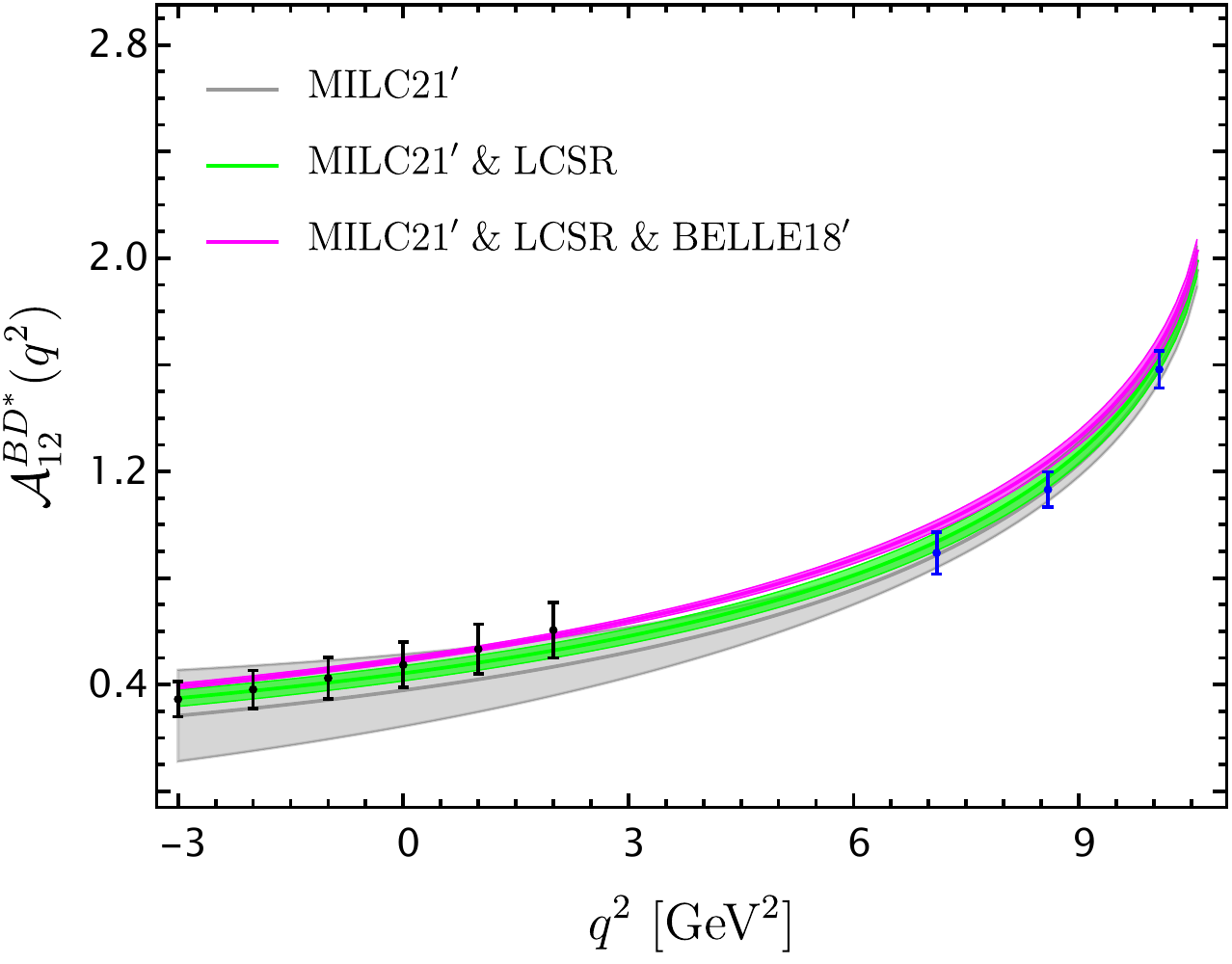}
\caption{Theory predictions for the momentum transfer dependence of the complete set of the exclusive
$\bar B \to D^{(\ast)} \, \ell \, \bar {\nu}_{\ell}$  form  factors in the entire kinematic region
from I) the BGL $z$-series fit against the  ``only lattice QCD" data points
\cite{MILC:2015uhg,Na:2015kha,FermilabLattice:2021cdg,McLean:2019qcx,Harrison:2021tol},
II) the simultaneous fit to both the lattice QCD results
\cite{MILC:2015uhg,Na:2015kha,FermilabLattice:2021cdg,McLean:2019qcx,Harrison:2021tol} and our LCSR predictions,
III) the combined numerical fit including further the available experimental data points
\cite{Belle:2015pkj,Belle:2018ezy,LHCb:2020cyw,LHCb:2020hpv}.}
\label{fig: q2 dependence of the B to D and Dstar form factors}
\end{center}
\end{figure*}

Inspecting the numerical features for the distinct classes of higher-order contributions
indicates that the newly determined NLL QCD corrections can reduce
the corresponding tree-level LP predictions  
by an amount of ${\cal O}(20 \, \%)$.
As displayed in Figure \ref{fig: Breakdown of the B to Dstar form factor A0},
the particular charm-quark mass dependent contribution $\hat{\xi}_{\|}^{\, m_c}$
in (\ref{effective form factor for xi_L}) appears to bring about the ${\cal O} (5 \, \%)$ enhancement
of the LP prediction of the form factor ${\cal A}_0$ with
$q^2 \in [-3.0, \, 2.0] \, {\rm GeV^2}$.
We further note that the effective form factor $\xi_{\|, \,  m_c}$ defined in (\ref{definition: NLP SCET longitudinal FF})
can only  result in the negligible  impact on the numerical prediction for ${\cal A}_0$
in the large recoil region due to the kinematical suppression of
the multiplication hard coefficient.
It remains important to remark that the yielding uncertainties of our  ${\rm NLL \oplus NLP}$  LCSR  predictions
arise, on the one hand, from the SCET computations of the  bottom-meson-to-vacuum correlation functions and,
on the other hand, from the extraction of the ground-state charmed meson contribution
with the parton-hadron duality ans\"{a}tz and the Borel transformation.
In order to determine the mean values and theoretical uncertainties of the $\bar B_q \to D_q^{\ast}$ form factors,
we employ the statistical procedure discussed in \cite{SentitemsuImsong:2014plu,Leljak:2021vte} by simultaneously scanning
the complete set of  input parameters (displayed in Table I of the Supplemental Material)
in the adopted intervals  with the prior distribution.
It is worthwhile mentioning further that  the systematic uncertainty due to the parton-hadron duality ans\"{a}tz
has been addressed in a wide range of  QCD computations (e.g., \cite{Chibisov:1996wf,Shifman:2000jv,Bigi:2001ys,Cata:2005zj,Beylich:2011aq,
Jamin:2011vd,Dingfelder:2016twb,Boito:2017cnp,Pich:2020gzz,Pich:2022tca}),
leading to the encouraging observation on the smallness of  duality violations
in the inclusive hadron production in $e^{+} \, e^{-}$ annihilations \cite{Pich:2020gzz},
in the hadronic decays of the $\tau$-lepton \cite{Pich:2022tca},
and in the semileptonic bottom-meson decays \cite{Dingfelder:2016twb}.
In an attempt to obtain more conservative predictions of our SCET sum rules,
we nevertheless increase the default intervals of  the sum-rule parameters $M^2$ and $s_0$
by a factor of two in the ultimate  error estimates.
As elaborated further in the Supplemental Material, one of the  principal benefits  from our LCSR analysis
consists in the yielding strong correlations between the LCSR data points at distinct $q^2$-values,
which  are expected to be particularly insensitive to the duality approximation.

In order  to extrapolate the LCSR predictions
for the $\bar B_q \to D_q^{\ast}$ form factors towards the large momentum transfer,
we will apply the BGL  parametrization \cite{Boyd:1995cf,Boyd:1995sq,Boyd:1997kz}
as widely employed  in the  form-factor determinations (see, for instance \cite{Bigi:2017njr,Bigi:2017jbd,Gambino:2019sif})
and then perform the binned $\chi^2$ fit of our LCSR predictions
at $q^2 \in \left \{-3.0, \, -2.0, \, -1.0,  \, 0.0, \, 1.0, \, 2.0  \right \} \, {\rm GeV}^2$,
in combination with the available lattice QCD results.
Moreover, we will implement the strong unitarity constraints
on the BGL coefficients $b_n^{i}$ by including  all the two-body
$\bar B_q^{(\ast)} \to D_q^{(\ast)}$ channels.
Consequently, we will  adopt the  HQET relations between these form factors
at ${\cal O}(\alpha_s, \, 1/m_b, \, 1/m_c^2)$,
allowing us to express the  $\bar B_q^{\ast} \to D_q^{(\ast)}$  form factors
in terms of the corresponding  $\bar B_q \to D_q^{(\ast)}$ form factors,
the NLP IW functions $\hat{\chi}_{2, \, 3}^{(q)}(\omega)$ and $\hat{\eta}^{(q)}(\omega)$ \cite{Neubert:1993mb},
and the ${\cal O} (1/m_c^2)$ IW functions $\hat{\ell}_{1,...,6}(\omega)$ \cite{Falk:1992wt}.
The determined intervals of the normalization and the slop parameters of these  IW functions
from \cite{Bordone:2019guc} will therefore be employed.

\begin{figure}
\begin{center}
\includegraphics[width=0.90 \columnwidth]{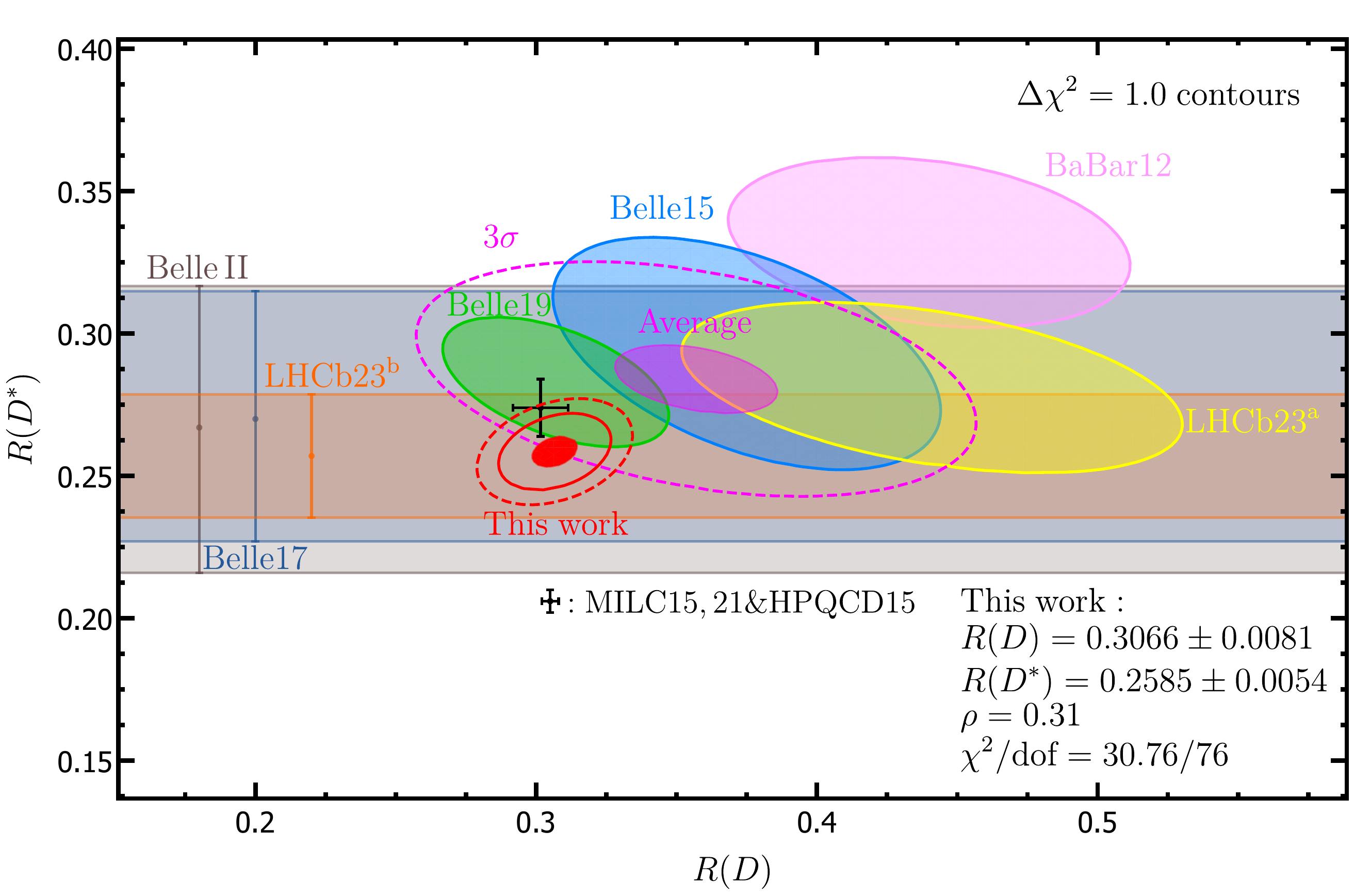}
\hspace{1.0 cm}
\caption{The correlated theory predictions for
${\cal R} ({D})$ and ${\cal R} ({D^{\ast}})$ from the combined BGL fits
against the lattice data points \cite{MILC:2015uhg,Na:2015kha,FermilabLattice:2021cdg}
and the ${\rm NLL \oplus NLP}$  LCSR predictions.
The available measurements from the BaBar \cite{BaBar:2012obs,BaBar:2013mob},
Belle \cite{Belle:2015qfa,Belle:2019rba,Belle:2016dyj},
LHCb \cite{LHCb:2023zxo,LHCb:2023cjr} and  Belle II Collaborations \cite{Belle-II-2023:RDstar} are also displayed.
The red solid and dashed  ellipses correspond to $95.45 \, \%$ and $99.73 \, \%$
confidence level contours, respectively.  }
\label{fig: numerics for RD and RDstar}
\end{center}
\end{figure}

Combining  the obtained LCSR results for  the  $\bar B_q \to D_q^{\ast}$ form factors
with I) the updated SCET sum rules for the $\bar B_q \to D_q$  form factors \cite{Gao:2021sav},
II) the lattice results for the  $\bar B \to D$ form factors
at $\omega =\left \{1.00, \, 1.08, \, 1.16  \right \}$
from the FNAL/MILC Collaboration \cite{MILC:2015uhg}
and the synthetic data points at $\omega =\left \{1.01, \, 1.06 \right \}$
from the HPQCD analysis \cite{Na:2015kha},
III) the unquenched lattice  results for the $\bar B \to D^{\ast}$ form factors
at  $\omega =\left \{1.03, \, 1.10, \, 1.17  \right \}$
from the FNAL/MILC Collaboration \cite{FermilabLattice:2021cdg},
IV) the lattice  determinations of the $\bar B_s  \to D_s$ form factors
at $\omega =\left \{1.0, \, 1.06, \, 1.12  \right \}$ from the HPQCD analysis \cite{McLean:2019qcx},
V) the  lattice computations of the $\bar B_s  \to D_s^{\ast}$ form factors
at $\omega =\left \{1.0, \, 1.04, \, 1.08 \right \}$ \cite{Harrison:2021tol},
we display in Table \ref{table: BGL fit results of the expansion coefficients} the resulting $z$-series coefficients
from the combined BGL fitting with the truncation $n=3$.
In order to better understand  the phenomenological  significance
of including  the  LCSR data points in the constrained BGL  fit,
we further carry out an alternative fit to the ``only lattice QCD" data points
of the  $\bar B \to D^{(\ast)}$ form factors \cite{MILC:2015uhg,Na:2015kha,FermilabLattice:2021cdg}
and the $\bar B_s \to D_s^{(\ast)}$ form factors \cite{McLean:2019qcx,Harrison:2021tol}.
It needs to be stressed that  our major objective is to investigate whether the inclusion of the large-recoil LCSR data points
in the BGL fit strategy allows us to increase effectively the precision of the yielding  form factors.
It is evident from Figure \ref{fig: q2 dependence of the B to D and Dstar form factors}
that taking into account the ${\rm NLL \oplus NLP}$ LCSR results will indeed be highly beneficial
for  pining down the uncertainties  of the $\bar B  \to D^{(\ast)}$ form factors
at small momentum transfer.
As the third fit model, we also carry out the statistical analysis
by taking into account  the experimental data points
for  $\bar B_{(s)} \to D_{(s)}^{(\ast)} \ell \bar {\nu}_{\ell}$
together with  the lattice QCD results and the LCSR  predictions.

We continue to present the correlated numerical predictions
for  ${\cal R}({D})$ and ${\cal R}({D^{\ast}})$
in Figure \ref{fig: numerics for RD and RDstar},
confronting with the available measurements from the BaBar \cite{BaBar:2012obs,BaBar:2013mob},
Belle \cite{Belle:2015qfa,Belle:2019rba,Belle:2016dyj},
LHCb   \cite{LHCb:2023zxo,LHCb:2023cjr} and Belle II Collaborations \cite{Belle-II-2023:RDstar}.
Importantly,  our  determinations of ${\cal R}({D})$ and ${\cal R}({D^{\ast}})$
by encompassing the  LCSR  results
in the numerical fit  deviate from the HFLAV-averaged measurements
by approximately $2.6 \, \sigma$, in contrast with the $3.3 \, \sigma$ discrepancy
between the arithmetic average  of the previous SM predictions \cite{Bigi:2016mdz,Bordone:2019vic,Gambino:2019sif}
and the state-of-the-art experimental average \cite{HFLAV:2022pwe}.
Subsequently, we carry out the simultaneous fit for  $|V_{cb}|$
and $|V_{ub}|$ by taking the determined $\bar B_{(s)} \to D_{(s)}^{(\ast)}$ form factors
 and the updated predictions of  the  $\bar B \to \pi$ form factors
in \cite{Cui:2022zwm}  in combination with the  lattice  data points
for the $\bar B_{(s)} \to D_{(s)}^{(\ast)}$ form factors
\cite{MILC:2015uhg,Na:2015kha,FermilabLattice:2021cdg,McLean:2019qcx,Harrison:2021tol}
and for the $\bar B \to \pi$ form factors \cite{Flynn:2015mha,FermilabLattice:2015mwy,FermilabLattice:2015cdh}
and the  experimental measurements
for $\bar B_{(s)} \to D_{(s)}^{(\ast)} \ell \bar {\nu}_{\ell}$ \cite{Belle:2015pkj,Belle:2018ezy,LHCb:2020cyw,LHCb:2020hpv}
and for $\bar B \to \pi \ell \bar {\nu}_{\ell}$ \cite{BaBar:2010efp,BaBar:2012thb,Belle:2010hep,Belle:2013hlo,Belle-II:2022imn}.
The yielding intervals of $|V_{cb}|$ and $|V_{ub}|$  are 
\begin{eqnarray}
\left \{ |V_{cb}|,  |V_{ub}|  \right \} =
\left \{ (39.64 \pm 0.63),   (3.71 \pm 0.13) \right \} \times 10^{-3},
\hspace{0.7 cm}
\label{numerics of Vub and Vcb}
\end{eqnarray}
which coincide with the previous exclusive extractions for $|V_{cb}|$ (for instance, \cite{Bigi:2017jbd,Grinstein:2017nlq,Gambino:2019sif,Bordone:2019vic,
Bordone:2019guc,Jaiswal:2017rve,Jaiswal:2020wer,Biswas:2022yvh,Martinelli:2021onb,Martinelli:2021myh,Martinelli:2022xir,Iguro:2020cpg})
and for $|V_{ub}|$ (e.g.,  \cite{Leljak:2021vte,Biswas:2021qyq,Biswas:2021cyd,Martinelli:2022tte}).
Unsurprisingly, the  obtained numerical predictions (\ref{numerics of Vub and Vcb})
remain to be in tension with  the world averages of the inclusive determinations
$|V_{cb}|= (42.2 \pm 0.8) \times 10^{-3}$ and
$|V_{ub}|= (4.13 \pm 0.12 ^{+0.13}_{-0.14} \pm 0.18) \times 10^{-3}$
\cite{ParticleDataGroup:2022pth} at the level of $2.5 \, \sigma$ and $1.5 \, \sigma$, respectively.
In an attempt to understand qualitatively the systematic uncertainties due to the  truncated  BGL expansions,
we increase the expansion order to $n=4$ and repeat the  numerical fit procedure for the form factors,
yielding the theory predictions for both the LFU ratios ${\cal R} (D^{(\ast)})$ and the CKM matrix elements
$|V_{cb}|$ and $|V_{ub}|$ only marginally different from the achieved results with $n=3$.

%
\section{Conclusions}
%

In conclusion, we have endeavored to accomplish for the first time the complete next-to-leading-order QCD computations
of the  $\bar B_{q} \to D_{q}^{(\ast)} \ell \bar {\nu}_{\ell}$ form factors at large recoil
and identified explicitly the unsuppressed charm-quark-mass dependent contributions
in the heavy quark expansion.
Taking into account the obtained  LCSR data points in the  BGL $z$-series fit
of  the  $\bar B_{q} \to D_{q}^{(\ast)} \ell \bar {\nu}_{\ell}$ form factors
enabled us to enhance significantly
the achieved accuracy of the large-recoil theory predictions
for the  $\bar B \to D^{(\ast)}$ form factors.
Implementing the determined LCSR results in the statistical analysis
turned out to be advantageous to  mitigate the combined ${\cal R}({D})$ and $ {\cal R}({D^{\ast}})$ tension
between the SM predictions and the  HFLAV-averaged measurements.
Our  computations of the heavy-to-heavy form factors near the maximal recoil
will be of notable importance for obtaining the yet higher precision predictions
of the  $\bar B_{q} \to D_{q}^{(\ast)} \ell \bar {\nu}_{\ell}$ decay observables,
when combined with the upcoming  lattice determinations of the bottom-meson distribution amplitudes.

%
\begin{acknowledgments}
\section*{Acknowledgements}

We are grateful to  Thomas Mannel, Zi-Hao Mi, Ru-Ying Tang, Alejandro Vaquero,
Chao Wang and Yan-Bing Wei for illuminating discussions.
Y.M.W. acknowledges support from the  National Natural Science Foundation of China  with
Grant No. 11735010 and 12075125, and the Natural Science Foundation of Tianjin
with Grant No. 19JCJQJC61100.

\end{acknowledgments}

\bibliographystyle{apsrev4-1}

\bibliography{References}

\newpage
\appendix
\begin{widetext}
\section{SUPPLEMENTAL MATERIAL}

\section{Analytic Expressions for the SCET Form Factors}

We first provide explicitly the conversion relations between the helicity form factors
$\{{\cal V}, \, {\cal A}_{0}, \, {\cal A}_{1}, \, {\cal A}_{12} \}$ and the conventional
form-factor basis   $\{V, \, A_0, \, A_1, \, A_{2} \}$  (see, for instance \cite{Beneke:2000wa})
\begin{eqnarray}
{\cal V} = {m_{B_q} \over m_{B_q} + m_{D_q^{\ast}}}  V,
\qquad
{\cal A}_0 =  {2 \, m_{D_q^{\ast}} \over n \cdot p} \, A_0,
\qquad
 {\cal A}_1 =   {m_B + m_{D_q^{\ast}} \over n \cdot p} \, A_1,
\qquad
{\cal A}_{12} =   {\cal A}_1  - {m_B - m_{D_q^{\ast}} \over m_B} \, A_2.
\end{eqnarray}
We then  collect the analytic expressions for the hard matching coefficients
in the ${\rm SCET_{I}}$ representations
of the semileptonic $\bar B_q \to D_q^{\ast}$ form factors at large hadronic recoil
\begin{eqnarray}
C_{f_{+}}^{\rm (A0)} &=& 1 + {\alpha_s \, C_F \over 4 \, \pi} \,
\left [ - 2 \ln^2 \left ({r \over \hat \mu}\right ) +  5 \,  \ln \left ({r \over \hat \mu}\right )
- 2 \, {\rm Li}_2 (1-r) - 3 \, \ln r - {\pi^2 \over 12} - 6 \right ] + {\cal O}(\alpha_s^2),
\nonumber \\
C_{f_{0}}^{\rm (A0)} &=& 1 + {\alpha_s \, C_F \over 4 \, \pi} \,
\left [ - 2 \ln^2 \left ({r \over \hat \mu}\right ) +  5 \,  \ln \left ({r \over \hat \mu}\right )
- 2 \, {\rm Li}_2 (1-r) - {3 - 5 r  \over 1 - r} \, \ln r - {\pi^2 \over 12} - 4 \right ] + {\cal O}(\alpha_s^2) \,,
\nonumber \\
C_{V}^{\rm (A0)} &=& 1 + {\alpha_s \, C_F \over 4 \, \pi} \,
\left [ - 2 \ln^2 \left ({r \over \hat \mu}\right ) +  5 \,  \ln \left ({r \over \hat \mu}\right )
- 2 \, {\rm Li}_2 (1-r) - {3 - 2\, r \over 1-r} \, \ln r - {\pi^2 \over 12} - 6 \right ] + {\cal O}(\alpha_s^2)\,,
\nonumber \\
C_{f_{+}}^{({\rm A1}, \, m_c)} &=& (r-1)+ {\cal O}(\alpha_s) \,,
\hspace{0.2 cm}
C_{f_{0}}^{({\rm A1}, \, m_c)} = (1-r)+ {\cal O}(\alpha_s) \,,
\hspace{0.2 cm}
C_{V}^{({\rm A1}, \, m_c)} = -1  + {\cal O}(\alpha_s)  \,,
\hspace{0.2 cm}
C_{A_1}^{({\rm A1}, \, m_c)} = 1  + {\cal O}(\alpha_s)   \,,
\nonumber \\
C_{f_{+}}^{\rm (B1)} &=& \left (-2  + {1 \over r} \right ) + {\cal O}(\alpha_s) \,,
\qquad
C_{f_{0}}^{\rm (B1)} =  \left (- {1 \over r} \right ) + {\cal O}(\alpha_s)  \,,
\qquad
C_{V}^{\rm (B1)} =  0 + {\cal O}(\alpha_s) \,,
\label{explicit results of the hard functions}
\end{eqnarray}
with the two dimensionless quantities $r=n \cdot p /m_b$ and $\hat \mu=\mu / m_b$.

We present  further the  effective Lagrangian densities ${\cal L}_{\xi m_c}^{(0)}$ and
${\cal L}_{\xi q, m_q}^{(2)}$  together with  the ${\rm SCET_I}$ weak current $O_{\|}^{\rm (A0)}$
entering the SCET correlation function $\Pi_{\nu, \, \|}^{\rm (A0)}$
\begin{eqnarray}
{\cal L}_{\xi m_c}^{(0)} &=&
m_c \, \bar \xi \left [i \slashed{D}_{\perp c}, \,  {1 \over i n \cdot D_c} \right ]  {\slashed{n} \over 2}  \xi
- m_c^2 \,  \bar \xi {1 \over  i n \cdot D_c}  {\slashed{n} \over 2} \xi \,,
\nonumber \\
{\cal L}_{\xi q, m_q}^{(2)} &=&
- m_q \, \left [ \left ( \bar \xi  W_c \right )  \, \left ( Y_s^{\dagger} q_s \right )
+ \left ( \bar q_s Y_s \right )  \, \left ( W_c^{\dagger} \xi \right )  \right ] \,,
\nonumber \\
O_{\|}^{\rm (A0)} &=&  \left ( \bar \xi W_c \right )  \gamma_5 \,  h_v \,,
\end{eqnarray}
which are evidently in demand for the construction of  the SCET sum rules (5) for the effective form factor $\xi_{\|}$
in the main text.
The sample Feynman diagrams for the one-loop computation of the correction function  $\Pi_{\nu, \, \|}^{\rm (A0)}$
with the aid of the perturbative factorization technique are collected  in (a), (b) and (c) in Figure \ref{fig: SCET Feynman diagrams}.
In particular, the three effective bottom-meson distribution amplitudes
$\phi_{B, \, {\rm eff}}^{-}$, $\phi_{B, \,  {\rm eff}}^{+, \, m_c}$ and $\phi_{B, \,  {\rm eff}}^{+, \, m_q}$
in the NLO sum rules  for the SCET form factor $\xi_{\|}$ can be  written as
\begin{eqnarray}
\phi_{B, \, {\rm eff}}^{-}(\omega^{\prime}) &=&
\theta(\omega^{\prime} - \omega_c) \,\phi_B^{-}(\omega^{\prime}- \omega_c)
+ {\alpha_s \, C_F \over 4 \, \pi} \,
\bigg \{  \int_0^{\infty} d \omega
\left [ \theta(\omega^{\prime} - \omega - \omega_c) \, \varrho_1(\omega, \omega^{\prime})
+ \theta(\omega + \omega_c -\omega^{\prime} ) \, \varrho_2(\omega, \omega^{\prime}) \right ] \,
{d \over d \omega} \, \phi_B^{-}(\omega)
\nonumber \\
&& + \, \int_0^{\infty} d \omega \,
\varrho_3(\omega, \omega^{\prime}) \, \phi_B^{-}(\omega)
- 4 \, \omega_c \left [ 3\, \ln {\mu \over m_c} + 2  \right ] \, {d \over d \omega^{\prime}}
\left [ \phi_B^{-}(\omega^{\prime}- \omega_c)  \, \theta(\omega^{\prime}- \omega_c) \right ] \bigg \}
+ {\cal O}(\alpha_s^2),
\nonumber \\
\phi_{B, \,  {\rm eff}}^{+, \, m_c}(\omega^{\prime}) &=&
- {\alpha_s \, C_F \over 4 \, \pi} \, \theta(\omega^{\prime} - \omega_c) \,
\int_0^{\infty} d \omega \, \left [  \theta(\omega^{\prime} - \omega - \omega_c) \, \varrho_4(\omega, \omega^{\prime})
+  \theta(\omega + \omega_c -\omega^{\prime} ) \, \varrho_5(\omega, \omega^{\prime}) \right ] \,
{d \over d \omega} \,  {\phi_B^{+}(\omega) \over \omega}  + {\cal O}(\alpha_s^2),
\nonumber \\
\phi_{B, \,  {\rm eff}}^{+, \, m_q}(\omega^{\prime}) &=&
{\alpha_s \, C_F \over 4 \, \pi} \, \theta(\omega^{\prime} - \omega_c) \,
\bigg \{  \int_0^{\infty} d \omega \,
\left [  \theta(\omega^{\prime} - \omega - \omega_c) \, \varrho_6(\omega, \omega^{\prime})
+  \theta(\omega + \omega_c -\omega^{\prime} ) \, \varrho_7(\omega, \omega^{\prime}) \right ] \,
{d \over d \omega} \,  {\phi_B^{+}(\omega) \over \omega}
\nonumber \\
&& + \int_0^{\infty} d \omega \,
\left [  \theta(\omega^{\prime} - \omega - \omega_c) \, \varrho_8(\omega, \omega^{\prime})
+  \theta(\omega + \omega_c -\omega^{\prime} ) \, \varrho_9(\omega, \omega^{\prime}) \right ] \,
 {\phi_B^{+}(\omega) \over \omega^2}  \bigg \}  + {\cal O}(\alpha_s^2),
\label{Results of the longtudinal effective B-meson LCDA}
\end{eqnarray}
where we have introduced the new coefficient functions $\varrho_i$ (with $i=1,..., 9$) for convenience
\begin{eqnarray}
\varrho_1(\omega, \omega^{\prime}) &=&
\ln {\omega^{\prime} - \omega- \omega_c \over \omega^{\prime} - \omega_c} \,
\left [ \left (1- {\omega_c  \over \omega^{\prime}} \right)^2
- 4 \, \ln {\omega_c \over \omega^{\prime} - \omega- \omega_c }
- 2 \, \ln {\mu^2 \over  n \cdot p \, \omega^{\prime} } -2  \right ] \,,
\nonumber \\
\varrho_2(\omega, \omega^{\prime}) &=&
\ln {\omega + \omega_c - \omega^{\prime}  \over \omega^{\prime} - \omega_c} \,
\left [ \left (1- {\omega_c  \over \omega^{\prime}} \right)^2
+ 4 \, \ln {\mu \over \omega^{\prime} - \omega_c}
+ 2 \, \ln {\omega^{\prime} \over n \cdot p } + 2 \right ]
+ 2 \, {\rm Li}_2  \left ( {\omega_c \over \omega_c - \omega^{\prime} } \right)
\nonumber \\
&& + \ln {\omega_c  \over \omega^{\prime} - \omega_c} \,
\left [ \left (1- {\omega_c  \over \omega^{\prime}} \right)^2
- 2 \, \ln {\omega_c \over  \omega^{\prime}}  - 2 \right ]
- \ln^2 {\mu^2 \over n \cdot p \, (\omega^{\prime} - \omega_c)}
- {\omega_c \over \omega^{\prime}}  - { 5 \, \pi^2 \over 6} \,,
\nonumber \\
\varrho_3(\omega, \omega^{\prime}) &=&
\theta(\omega^{\prime} - \omega - \omega_c) \, {1 \over \omega + \omega_c -\omega^{\prime}} \,
\left [ 4 \, \ln {\omega^{\prime} - \omega \over \omega_c}
- \left ( {\omega^{\prime} - \omega - \omega_c \over \omega^{\prime} - \omega } \right )^2  \right ]
+  \theta(\omega + \omega_c -\omega^{\prime}) \, {2 \over \omega} \,,
\nonumber \\
\varrho_4(\omega, \omega^{\prime}) &=& { (\omega^{\prime} - \omega_c)^2 \over \omega^{\prime}} \,
\ln { \omega^{\prime} - \omega - \omega_c \over \omega^{\prime} - \omega_c}
- \omega^{\prime} \, \ln {\omega^{\prime} - \omega \over \omega^{\prime}}
- {\omega\, \omega_c \over  \omega^{\prime} - \omega } \,,
\nonumber \\
\varrho_5(\omega, \omega^{\prime}) &=& { (\omega^{\prime} - \omega_c)^2 \over \omega^{\prime}} \,
\ln { (\omega + \omega_c - \omega^{\prime}) \, \omega_c \over  (\omega^{\prime} - \omega_c)^2 }
- \omega^{\prime} \, \ln {\omega_c \over \omega^{\prime}} \,,
\nonumber \\
\varrho_6(\omega, \omega^{\prime}) &=&
\left[ 2 \, \omega^{\prime} + \omega_c \, \left (1 + {\omega_c  \over \omega^{\prime} } \right )  \right ] \,
\ln { \omega^{\prime} - \omega_c \over \omega^{\prime} - \omega_c  - \omega} \,,
\nonumber \\
\varrho_7(\omega, \omega^{\prime}) &=&
\left[ 2 \, \omega^{\prime} + \omega_c \, \left (1 + {\omega_c  \over \omega^{\prime} } \right )  \right ] \,
\ln { (\omega^{\prime} - \omega_c)^2 \over \omega_c \, (\omega + \omega_c  - \omega^{\prime}) }
- \left ( 2 \, \ln {\mu^2 \over m_c^2} + 5 \right ) \, \omega^{\prime} + \omega_c   \,,
\nonumber \\
\varrho_8(\omega, \omega^{\prime}) &=&
2 \, \omega^{\prime} \, \left [  \ln { \omega^{\prime}  - \omega \over \omega^{\prime} }
+ 2 \,  \ln {\omega^{\prime}  - \omega_c \over \omega^{\prime}  - \omega_c  - \omega} \right ]
+ {\omega \, \omega_c \over \omega - \omega^{\prime}} \,,
\nonumber \\
\varrho_9(\omega, \omega^{\prime}) &=& - \omega^{\prime} \, \left [  4 \, \ln {\mu \over \omega^{\prime}  - \omega_c }
+ 2 \, \ln {\omega^{\prime} \over n \cdot p }  + 5 \right ] + \omega_c  \,.
\end{eqnarray}

\begin{figure}[tp]
\begin{center}
\includegraphics[width=0.75 \columnwidth]{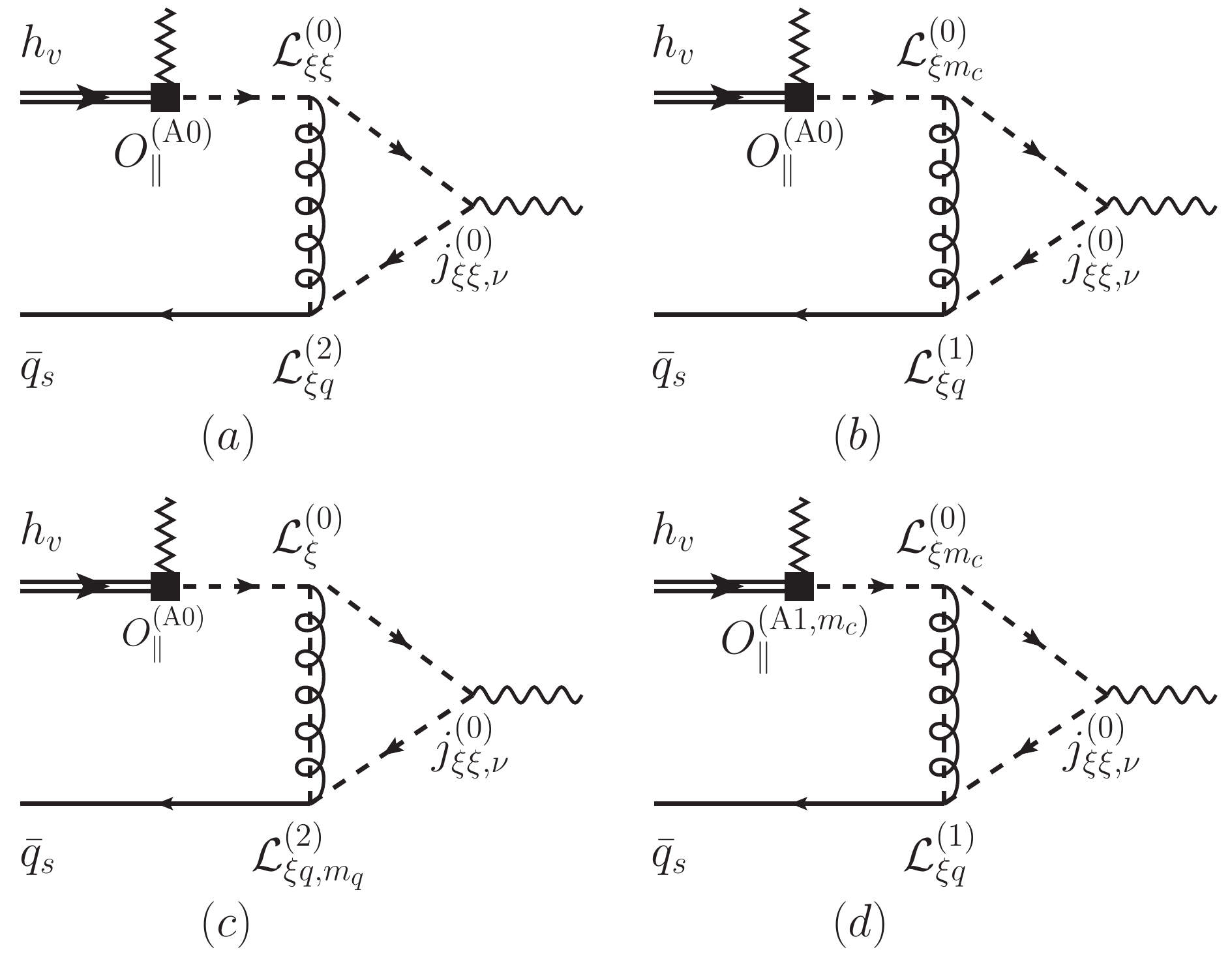}
\caption{Sample diagrams for the two bottom-meson-to-vacuum correlation functions
$\Pi_{\nu, \, \|}^{\rm (A0)}$ and $\Pi_{\nu, \, \|}^{{\rm (A1)}, m_c}$ in ${\rm SCET_I}$.}
\label{fig: SCET Feynman diagrams}
\end{center}
\end{figure}

Along the same vein,  we can establish the desired LCSR for the effective form factor $\xi_{\|, \,  m_c}$ by employing
the bottom-meson-to-vacuum correction function
\begin{eqnarray}
&& \Pi_{\nu, \, \|}^{({\rm A1}, \, m_c)}
=   \int d^4 x \, e^{i p \cdot x} \, \int d^4 y  \, \int d^4 z  \,
\langle  0 | {\rm T} \{j_{\xi \xi, \| \nu}^{(0)}(x), \, i {\cal L}_{\xi q}^{(1)}(y),
\, i {\cal L}_{\xi m_c}^{(0)}(z), \, O_{\|}^{({\rm A1}, \, m_c)}(0) \} | \bar B_v \rangle,
\label{definition: the correlation function of A1-mc}
\end{eqnarray}
where the manifest form of the ${\rm A1}$-type  ${\rm SCET}_{\rm I}$ current is given by
\begin{eqnarray}
O_{\|}^{({\rm A1}, \, m_c)}(0) =  (\bar \xi W_c) \,  {\slashed{n} \over 2}
{m_c  \over - i n \cdot \overleftarrow{D}_c} \gamma_5 \,   h_v  \,.
\end{eqnarray}
It turns out that only the diagram (d) in Figure \ref{fig: SCET Feynman diagrams}
can yield the non-vanishing contribution to the correlation function
(\ref{definition: the correlation function of A1-mc}) at ${\cal O}(\alpha_s)$.
Equating the obtained partonic representation with the counterpart  hadronic
dispersion relation allows us to derive the new sum rules
for  $\xi_{\|, \,  m_c}$
\begin{eqnarray}
\xi_{\|, \,  m_c} &=& 2 \, \frac{{\cal F}_{B_q}(\mu)}{f_{D_q^{\ast}, \|}}
{m_{B_q} m_{D_q^{\ast}} \over (n \cdot p)^2}  \int_0^{\omega_s} d \omega^{\prime}
{\rm exp} \left [  { m_{D_q^{\ast}} ^2  - n \cdot p \, \omega^{\prime}   \over n \cdot p \, \omega_M}  \right ]
\, \left [{ \omega_c  \over  \omega^{\prime}} \, \phi_{B, \,  {\rm eff}}^{+, \, m_c}(\omega^{\prime}) \right ],
\end{eqnarray}
with $\omega_c = m_c^2/n \cdot p$.
We mention in passing that the default power counting rules for $\omega_s$, $\omega_M$ and $\omega_c$ are
\begin{eqnarray}
\omega_{c} \sim  \omega_{s} \sim  {\cal O}(\Lambda_{\rm QCD}), \qquad
(\omega_s -\omega_c) \sim  {\cal O}(\Lambda_{\rm QCD}^{3/2} / m_b^{1/2}),
\qquad
\omega_M \sim {\cal O}(\Lambda_{\rm QCD}^{3/2} / m_b^{1/2}),
\end{eqnarray}
which differ from the conventional counting schemes for the exclusive charmless bottom-hadron decays
\cite{Wang:2015vgv,Wang:2015ndk,Wang:2016qii,Wang:2018wfj,Shen:2020hfq}.

Now we turn to derive the  SCET  sum rules  for  the ${\rm B1}$-type longitudinal  form factor
$\Xi_{\|}$ by investigating the bottom-meson-to-vacuum correlation function
\begin{eqnarray}
\Pi_{\nu, \, \|}^{({\rm B1})}
= {n \cdot p \over 2 \pi}  \int d^4 x \, e^{i p \cdot x} \, \int dr \, e^{-i n \cdot p \, \tau \, r}  \, \int d^4 y  \,
\, \langle  0 | {\rm T} \{j_{\xi \xi, \| \nu}^{(0)}(x), \, i {\cal L}_{\xi q}^{(1)}(y),
\,\,  O_{\|}^{({\rm B1})}(r\, n) \} | \bar B_v \rangle,
\label{definition: the correlation function of B1}
\end{eqnarray}
where the non-local  effective current $O_{\|}^{({\rm B1})}$ reads
\begin{eqnarray}
O_{\|}^{({\rm B1})} =  (\bar \xi W_c)(0) \, \gamma_5 \,
(W_c^{\dag} \, i \slashed{D}_{c \perp} \, W_c)(r n) \,   h_v(0) \,.
\end{eqnarray}
It is then straightforward to construct the one-loop LCSR for
the non-local form factor $\Xi_{\|}$ with the HQET bottom-meson distribution amplitude
\cite{Gao:2019lta}
\begin{eqnarray}
 \Xi_{\|} = -{\alpha_s \, C_F \over \pi} \, \frac{{\cal F}_{B_q}(\mu)}{f_{D_q^{\ast}, \|}}
{m_{B_q} m_{D_q^{\ast}} \over n \cdot p \, m_b} \,
\bar \tau  \, \theta(\tau) \, \theta \left (\bar \tau - {\omega_c \over \omega_s} \right ) \,
  \int_{\omega_c/\bar \tau}^{\omega_s} d \omega^{\prime} \,
{\rm exp} \left [  { m_{D_q^{\ast}} ^2  - n \cdot p \, \omega^{\prime}   \over n \cdot p \, \omega_M}  \right ]
\int_{\omega^{\prime}  - \omega_c/\bar \tau}^{+\infty} \, {d \omega \over \omega} \, \phi_B^{+}(\omega),
\end{eqnarray}
with  $\bar \tau = 1- \tau$.
Adopting the factorization scale of order $\sqrt{m_b\, \Lambda_{\rm QCD}}$,
we continue to perform  the next-to-leading-logarithmic (NLL) resummation
for the  enhanced logarithms in the final expressions
of the considered  QCD form factors by employing the standard renormalization-group (RG) formalism
(see \cite{Cui:2022zwm,Gao:2021sav} for further details).

\begin{table}[tp]
\centering
\renewcommand{\arraystretch}{1.5}
\resizebox{0.8 \textwidth}{!}{
\begin{tabular}{|c| c| c| c| c|}
    \hline
    \hline
    Parameter                                   & Value/Interval                 & Unit     & Prior               & Source/Comments\\
    \hline
    \multicolumn{5}{|c|}{Quark-Gluon Coupling and Quark Masses}\\
    \hline
    $\alpha_s^{(5)}(m_Z)$                       &  $[0.1170, \, 0.1188]$         & ---      & --- &  \cite{ParticleDataGroup:2022pth}                \\
    $\overline{m}_b(\overline{m}_b)$            & $[4.192,  \,   4.214]$         & GeV      & gaussian $@$ $68 \%$  & \cite{ParticleDataGroup:2022pth}  \\
    $m^{\rm PS}_b({\rm 2 \, GeV})$              & $[4.493, \,  4.545]$           & GeV      & gaussian $@$ $68 \%$ & \cite{Beneke:2014pta} \\
    $\overline{m}_c(\overline{m}_c)$            & $[1.265,  \,  1.291]$          & GeV      & gaussian $@$ $68 \%$ & \cite{FlavourLatticeAveragingGroupFLAG:2021npn}  \\
    $\overline{m}_s({\rm 2 \, GeV})$              & $[92.5,  \,  93.7]$          & MeV      & gaussian $@$ $68 \%$ &  \cite{ParticleDataGroup:2022pth}\\
    $\overline{m}_u({\rm 2 \, GeV})$              & $[2.12,  \,  2.28]$          & MeV      & gaussian $@$ $68 \%$ &  \cite{ParticleDataGroup:2022pth}\\
    $\overline{m_d}({\rm 2 \, GeV})$              & $[4.64,  \,  4.74]$          & MeV      & gaussian $@$ $68 \%$ &  \cite{ParticleDataGroup:2022pth}\\
    \hline
    \multicolumn{5}{|c|}{Hadron Masses}\\
    \hline
    $m_B$                                       &  5279.66           & MeV     & ---                 &  \cite{ParticleDataGroup:2022pth}\\
    $m_{B_s}$                                   &  5366.92           & MeV     & ---                 &  \cite{ParticleDataGroup:2022pth}\\
    $m_D$                                       &  1869.66           & MeV     & ---                 &  \cite{ParticleDataGroup:2022pth}\\
    $m_{D^*}$                                   &  2010.26           & MeV     & ---                 &  \cite{ParticleDataGroup:2022pth}\\
    $m_{D_s}$                                   &  1968.35           & MeV     & ---                 &  \cite{ParticleDataGroup:2022pth}\\
    $m_{D_s^*}$                                 &  2112.2            & MeV     & ---                 &  \cite{ParticleDataGroup:2022pth}\\
    \hline
    \multicolumn{5}{|c|}{Decay Constants}\\
    \hline
    $f_{B_d} |_{N_f=2+1+1}$                    & $[0.1887,  \,  0.1913]$   & GeV & gaussian $@$ $68 \%$ & \cite{FlavourLatticeAveragingGroupFLAG:2021npn}    \\
    $f_{B_s}|_{N_f=2+1+1}$                     & $[0.2290,  \,  0.2316]$   & GeV & gaussian $@$ $68 \%$ & \cite{FlavourLatticeAveragingGroupFLAG:2021npn}    \\
    $f_D |_{N_f=2+1+1}$                        & $[0.2113,  \,  0.2127]$   & GeV & gaussian $@$ $68  \%$ & \cite{FlavourLatticeAveragingGroupFLAG:2021npn}   \\
    $f_{D_s} |_{N_f=2+1+1}$                    & $[0.2494,  \,  0.2504]$   & GeV & gaussian $@$ $68 \%$ & \cite{FlavourLatticeAveragingGroupFLAG:2021npn}   \\
    $f_{D^{\ast}, \|} $                        & $[0.210,   \, 0.245]$     & GeV & gaussian $@$ $68  \%$  & \cite{Pullin:2021ebn}\\
    $f_{D^{\ast}, \perp}(\nu)$                 & $[0.186,   \, 0.218]$     & GeV & gaussian $@$ $68  \%$  & \cite{Pullin:2021ebn}\\
    $f_{D_s^{\ast}, \|}$                       & $[0.260,   \, 0.298]$     & GeV & gaussian $@$ $68  \%$  & \cite{Pullin:2021ebn}\\
    $f_{D_s^{\ast}, \perp}(\nu)$               & $[0.239,   \, 0.272]$   & GeV & gaussian $@$ $68  \%$    & \cite{Pullin:2021ebn}\\
    \hline
    \multicolumn{5}{|c|}{Shape Parameters of the Bottom-Meson LCDAs}\\
    \hline
    $\lambda_{B_d}(\mu_0)$                                  & $[200,   \, 500]$          & MeV    & uniform $@$ $100 \%$   & \cite{Beneke:2020fot,Shen:2020hfq} \\
    $\lambda_{B_s}(\mu_0)$                                  & $[250,   \, 550]$          & MeV    & uniform $@$ $100 \%$    & \cite{Beneke:2020fot,Shen:2020hfq} \\
    $\left \{\hat{\sigma}_1(\mu_0), \, \hat{\sigma}_2(\mu_0) \right \}$
                                                            & $\left [\{-0.7, \,  -6.0\},  \,\, \{0.7,  \, 6.0\} \right ]$
                                                                                          & ---    & uniform $@$ $100 \%$
    &\cite{Beneke:2020fot,Shen:2020hfq} \\
    $\lambda_E^2(\mu_0) / \lambda_H^2(\mu_0)$               & $[0.40,   \, 0.60]$         & ---    & uniform $@$ $100 \%$    &\cite{Beneke:2018wjp} \\
    $2 \, \lambda_E^2(\mu_0) + \lambda_H^2(\mu_0)$          & $[100,   \, 400]$           & MeV    & uniform $@$ $100 \%$    & \cite{Beneke:2018wjp} \\
    \hline
    \multicolumn{5}{|c|}{Sum Rule Parameters and Scales}\\
    \hline
    $\mu$           &  $[1.0,   \, 2.0]$           & GeV     & uniform $@$ $100 \%$                 &  \cite{Cui:2022zwm} \\
    $\mu_h$         &  $[m_b/2, \, 2\, m_b]$       & GeV     & $\ln(\mu_h)$ uniform $@$ $100\%$     &  \cite{Cui:2022zwm} \\
    $M^2$           &  $[3.5,   \, 5.5]$           & GeV$^2$ & uniform $@$ $100 \%$                  &  \cite{Wang:2017jow,Khodjamirian:2020mlb,Gao:2021sav}  \\
    $s^{D}_0 \, (s^{D^{\ast}}_0)$       &  $[6.5,   \, 7.5]$           & GeV$^2$ & uniform $@$ $100 \%$                 & \cite{Wang:2017jow,Khodjamirian:2020mlb,Gao:2021sav}  \\
    $s^{D_s}_0 \, (s^{D_s^{\ast}}_0)$     &  $[7.0,   \, 8.0]$           & GeV$^2$ & uniform $@$ $100 \%$                 & \cite{Duplancic:2015zna,Li:2020rcg} \\
    \hline
    \hline
\end{tabular}
}
\caption{Numerical values of the input parameters employed in the LCSR determinations
of the exclusive $\bar B_{q} \to D_{q}^{(\ast)} \ell \bar {\nu}_{\ell}$  form factors.
The quoted transverse $D_q^{\ast}$ decay constants are evaluated
at the  renormalization scale  $\nu=1.67 \, {\rm GeV}$ \cite{Pullin:2021ebn}.
The full prior distribution is determined by the product of the uncorrelated individual priors,
which are either uniform or Guassian distributed. The uniform priors cover the listed intervals
with $100 \, \%$ probability, while the Gaussian ones adopt the listed intervals such that
the central values correspond to the modes and the intervals contain $68 \, \%$ of accumulated probability \cite{SentitemsuImsong:2014plu,Leljak:2021vte}. }
\label{tab:input paraeters}
\end{table}

Likewise, we can construct the SCET sum rules for the  transverse form factors
$\xi_{\perp}$, $\xi_{\perp, \,  m_c}$ and $\Xi_{\perp}$
by adopting three additional correlation functions
$\Pi_{\mu \nu \rho, \, \perp}^{\rm (A0)}$, $\Pi_{\mu \nu \rho, \, \perp}^{({\rm  A1}, \, m_c)}$
and $\Pi_{\mu \nu \rho, \, \perp}^{\rm (B1)}$,
which can be  obtained from the counterpart longitudinal correlation functions
$\Pi_{\nu, \, \|}^{\rm (A0)}$, $\Pi_{\nu, \, \|}^{({\rm A1}, \, m_c)}$ and
$\Pi_{\nu, \, \|}^{({\rm B1})}$
with the appropriate replacement rules (see \cite{Gao:2019lta}).
\begin{eqnarray}
j_{\xi q, \, \| \nu}^{(0, 2)} \to j_{\xi q, \, \perp \nu \rho}^{(0, 2)}  \,,
\qquad
O_{\|}^{(\rm A0)} \to O_{\perp, \, \mu}^{(\rm A0)} \,,
\qquad
O_{\|}^{({\rm A1}, \, m_c)} \to O_{\perp, \, \mu}^{({\rm A1}, \, m_c)} \,,
\qquad
O_{\|}^{(\rm B1)} \to O_{\perp, \, \mu}^{(\rm B1)} \,.
\end{eqnarray}
The interpolating currents $j_{\xi q, \, \perp \nu \rho}^{(0, 2)}$ and the necessary SCET weak currents are given by
\begin{eqnarray}
j_{\xi q, \, \perp \nu \rho}^{(0)}
&=&  \bar \xi \, {\slashed {n} \over 2} \, \gamma_{\perp \rho} \, \xi \, \bar n_{\nu} \,,
\qquad
j_{\xi q, \, \perp \nu \rho}^{(2)}
=  \left ( \bar \xi W_c \, \, {\slashed {n} \over 2} \, \gamma_{\perp \rho} \,\,  Y_s^{\dag} q
+ \bar q  Y_s \, \, {\slashed {n} \over 2} \, \gamma_{\perp \rho} \,\, W_c^{\dag} \xi \right ) \bar n_{\nu}  \,,
\nonumber \\
O_{\perp, \, \mu}^{(\rm A0)}
&=& \left ( \bar \xi W_c \right )  \gamma_5 \, \gamma_{\perp \mu} \,  h_v  \,,
\qquad
O_{\perp, \, \mu}^{({\rm A1}, \, m_c)} = m_c \,
(\bar \xi W_c) \,  {\slashed{n} \over 2}
{1 \over - i n \cdot \overleftarrow{D}_c} \gamma_5 \, \gamma_{\perp \mu}  \,   h_v \,,
\nonumber \\
O_{\perp, \, \mu}^{(\rm B1)}  &=&
 (\bar \xi W_c)(0) \, \gamma_5 \,  \gamma_{\perp \mu}   \,
(W_c^{\dag} \, i \slashed{D}_{c \perp} \, W_c)(r n) \,   h_v(0) \,.
\end{eqnarray}
Interestingly,  the effective form factor $\xi_{\perp, \, m_c}$
vanishes at one loop in the four-dimensional space-time.
The resulting SCET sum rules for the effective transverse form factors
$\xi_{\perp}$ and $\Xi_{\perp}$ at ${\cal O}(\alpha_s)$ can be explicitly written as
\begin{eqnarray}
\xi_{\perp} &=&  \frac{{\cal F}_{B_q}(\mu)}{f_{D_q^{\ast}, \perp}} \,
{m_{B_q}  \over n \cdot p} \, \int_0^{\omega_s} d \omega^{\prime} \,
{\rm exp} \left [  {m_{D_q^{\ast}} ^2 - n \cdot p \, \omega^{\prime}    \over n \cdot p \, \omega_M}  \right ]
\, \left \{ \left [ \phi_{B, \, {\rm eff}}^{-}(\omega^{\prime}) + \Delta \phi_{B, \, {\rm eff}}^{-}(\omega^{\prime})  \right ]
+  { m_q \over \omega^{\prime}} \, \left [ \phi_{B, \,  {\rm eff}}^{+, \, m_q}(\omega^{\prime})
+ \Delta \phi_{B, \,  {\rm eff}}^{+, \, m_q}(\omega^{\prime})  \right ]
\right \},
\nonumber \\
\nonumber \\
\Xi_{\perp} &=& -{\alpha_s \, C_F \over 2 \, \pi} \, \frac{{\cal F}_{B_q}(\mu)}{f_{D_q^{\ast}, \perp}}
{m_{B_q}  \over  m_b} \,
\bar \tau  \, \theta (\tau) \, \theta \left (\bar \tau - {\omega_c \over \omega_s} \right) \,
\int_{\omega_c/\bar \tau}^{\omega_s} d \omega^{\prime}
\,  {\rm exp} \left [  { m_{D_q^{\ast}} ^2  - n \cdot p \, \omega^{\prime}   \over n \cdot p \, \omega_M}  \right ] \,
\int_{\omega^{\prime}  - \omega_c/\bar \tau}^{+\infty} \, {d \omega \over \omega} \, \phi_B^{+}(\omega),
\label{SCET sum rules for the effective transverse form factors}
\end{eqnarray}
where we have introduced two additional coefficient functions $\Delta \phi_{B, \, {\rm eff}}^{-}$
and $\Delta \phi_{B, \,  {\rm eff}}^{+, \, m_q}$ for brevity
\begin{eqnarray}
\Delta \phi_{B, \, {\rm eff}}^{-}(\omega^{\prime})
&=& {\alpha_s \, C_F \over 4 \, \pi} \, \theta(\omega^{\prime} - \omega_c) \,
\int_0^{\infty} d \omega \, \bigg  \{ \theta(\omega^{\prime} - \omega - \omega_c) \,
\left ( 1 - {\omega_c^2 \over \omega^{\prime 2}}  \right )  \,
\ln {\omega^{\prime} - \omega - \omega_c \over \omega^{\prime} - \omega_c}
\nonumber \\
&& + \, \theta(\omega + \omega_c - \omega^{\prime}) \,
\left [ \ln {\mu^2 \over m_c^2}  + \left ( 1 - {\omega_c^2 \over \omega^{\prime 2}}  \right )  \,
\ln {\omega_c \, (\omega + \omega_c - \omega^{\prime})  \over (\omega^{\prime} - \omega_c)^2 }
+ {\omega_c \over \omega^{\prime}}  \right ]  \bigg \} \,
{d \over d \omega} \, \phi_B^{-}(\omega)  + {\cal O}(\alpha_s^2),
\nonumber \\
\Delta \phi_{B, \,  {\rm eff}}^{+, \, m_q}(\omega^{\prime})
&=&  {\alpha_s \, C_F \over 4 \, \pi} \, \theta(\omega^{\prime} - \omega_c) \,
\int_0^{\infty} d \omega \,
\bigg \{ \theta(\omega^{\prime} - \omega - \omega_c) \,
\left [  {\omega_c \, (\omega^{\prime} - \omega_c) \over \omega^{\prime}}  \,
\ln { \omega^{\prime} - \omega_c  \over  \omega^{\prime} - \omega_c - \omega } \,
{d \over d \omega} \, {\phi_B^{+}(\omega) \over \omega}
+ {\omega \, \omega_c \over \omega^{\prime} - \omega} \,
{\phi_B^{+}(\omega)  \over \omega^2} \right ]
\nonumber \\
&& + \, \theta(\omega + \omega_c - \omega^{\prime}) \,
\left [  {\omega_c \, (\omega^{\prime} - \omega_c) \over \omega^{\prime}}  \,
\ln { (\omega^{\prime} - \omega_c)^2 \over  \omega_c \, (\omega + \omega_c - \omega^{\prime}) } \,
{d \over d \omega} \, {\phi_B^{+}(\omega) \over \omega}
+ {\omega^{\prime} - \omega_c  \over \omega} \, {d \over d \omega} \, \phi_B^{+}(\omega) \right ]  \bigg \}
 + {\cal O}(\alpha_s^2)  \,.
\end{eqnarray}

\section{Detailed Numerical Results for the Exclusive Bottom-Meson Decay Form Factors}

We summarize explicitly the numerical values of all input parameters employed in our LCSR analysis
and their prior density functions in Table  \ref{tab:input paraeters}, including further their sources.
As already discussed in the main text,  the statistical procedure previously discussed in \cite{SentitemsuImsong:2014plu,Leljak:2021vte}
is then implemented to determine the mean values and  theoretical uncertainties for the form factors of our interest.
In order to facilitate the future phenomenological explorations,
we  collect the yielding numerical results of the semileptonic $\bar B_{(s)} \to D_{(s)}^{(\ast)}$ decay  form factors
as well as their correlation matrix  at the six representative kinematic points
$q^2 \in \left \{-3.0, \, -2.0, \, -1.0,  \, 0.0, \, 1.0, \, 2.0  \right \} \, {\rm GeV}^2$
from our  ${\rm NLL \oplus NLP}$  LCSR computations in the ancillary file {\tt InputLCSR.txt}
attached to the arXiv preprint version of this letter.
Our numerical explorations indicate that the dominating theory uncertainties of the current LCSR computations
for the exclusive $b \to c \ell \bar \nu_{\ell}$ form factors arise from the  variations of non-perturbative shape parameters
dictating the two-particle bottom-meson distribution amplitudes in HQET.
The most prominent feature of our  LCSR predictions consists in the strong correlations of the obtained data points
at the different $q^2$-values, which can be attributed to the universal hadronic inputs appearing in the analytical sum rules
for the complete set of the  $\bar B_{q} \to D_{q}^{(\ast)}$ form factors.
This essential pattern is in sharp contrast with the more conventional sum rules for the heavy-to-light $B$-meson form factors,
employing the light-meson distribution amplitudes in QCD \cite{Belyaev:1993wp,Ali:1993vd,Ball:1997rj,Duplancic:2008ix,Khodjamirian:2011ub}.
It is then naturally expected that  the obtained correlation matrix for the LCSR data points would be  insensitive to
the parton-hadron duality violation contributions (see \cite{Chibisov:1996wf,Shifman:2000jv,Bigi:2001ys,Cata:2005zj,Beylich:2011aq,
Jamin:2011vd,Dingfelder:2016twb,Boito:2017cnp,Pich:2020gzz,Pich:2022tca} for further discussions in different contexts).
On the phenomenological aspect, the duality violation contribution in the LCSR framework can be  probed
by adopting the different alternatives for the duality ans\"{a}tz with the inclusion of
the radial excited or continuum states and with the increased effective threshold  \cite{Colangelo:2000dp}.

In addition, we provide manifestly the correlation matrices of the obtained BGL expansion coefficients
for the $\bar B_{(s)} \to D_{(s)}^{(\ast)}$ form factors with the three different fitting strategies
(as discussed in the main text)  in the ancillary file {\tt BGLCoeff.txt}
attached to the arXiv preprint version  for completeness.
Having at our disposal the correlated $z$-series coefficients, we are then prepared to arrive at
the major phenomenological predictions for the considered form factors and the gold-plated LFU quantities
as presented in Figure 2 and 3 of the main text.

\end{widetext}

\end{document}